\documentclass[conference]{IEEEtran}
\usepackage{multirow}

\usepackage{longtable}
\newcommand{\tabincell}[2]{\begin{tabular}{@{}#1@{}}#2\end{tabular}}
\ifCLASSINFOpdf
\else
\fi
\usepackage{array,longtable}
\usepackage{amsmath}
\usepackage{graphicx}
\usepackage[monochrome]{xcolor}
\newcommand{\tim}[1]{\color{blue}#1\color{black}}

\begin{document}
%
\title{\textcolor{red}{PUF Based Encryption Solution : Barrel shifter Physical Unclonable Functions}}
\title{\tim{Barrel Shifter Physical Unclonable Function Based Encryption}}

\author{\IEEEauthorblockN{Yunxi Guo}
\IEEEauthorblockA{Department of Electrical and\\ Computer Engineering\\
Iowa State University\\
Ames, IA 50011\\
Email: yunxig@iastate.edu}
\and
\IEEEauthorblockN{Timothy Dee}
\IEEEauthorblockA{Department of Electrical and\\ Computer Engineering\\
Iowa State University\\
Ames, IA 50011\\
Email: timdee@iastate.edu}
\and
\IEEEauthorblockN{Akhilesh Tyagi}
\IEEEauthorblockA{Department of Electrical and\\ Computer Engineering\\
Iowa State University\\
Ames, IA 50011\\
Email: tyagi@iastate.edu}
}

\maketitle

\begin{abstract}
\tim{
Physical Unclonable Functions (PUFs) are circuits designed to extract physical randomness from the underlying circuit.
This randomness depends on the manufacturing process.
It differs for each device enabling chip-level authentication and key generation \cite{suh2007physical} applications.
We present a protocol utilizing a PUF for secure data transmission. 
Parties each have a PUF used for encryption and decryption;
this is facilitated by constraining the PUF to be commutative.
This framework is evaluated with a primitive permutation network - a barrel shifter \cite{hashmi2010efficient}.
Physical randomness is derived from the delay of different shift paths. 
Barrel shifter (BS) PUF captures the delay of different shift paths.
This delay is entangled with message bits before they are sent across an insecure channel.
BS-PUF is implemented using transmission gates;
their characteristics ensure same-chip reproducibility, a necessary property of PUFs.
Post-layout simulations of a common centroid layout \cite{ma2007analog} 8-level barrel shifter in 0.13 $\mu m$ technology assess uniqueness, stability and randomness properties.
BS-PUFs pass all selected NIST statistical randomness tests \cite{rukhin2010statistical}.
Stability similar to Ring Oscillator (RO) PUFs under environment variation is shown. 
Logistic regression of $100,000$ plaintext-ciphertext pairs (PCPs) failed to successfully model BS-PUF behavior.
}

\end{abstract}

\IEEEpeerreviewmaketitle

\section{Introduction}
Encryption/decryption algorithms form the backbone of modern public key infrastructure which supports 
a broad set of activities such as e-commerce and digital currency. Mathematical cryptosystems
such as RSA \cite{takagi1998fast} can take millions of clock cycles. Even symmetric encryption/decryption
through AES takes 10-20 clock cycles.
Moreover, even though their security is
 predicated on
a hard mathematical problem such as prime number factoring, a mathematical model exists for an adversary \cite{boneh1999twenty}.
Physical unclonable functions (PUFs) source physical randomness of a silicon foundry with a potential appeal of unmodelable, physical functions. They have been used to
generate unique physical identities, and to seed key generation. Such PUFs offer both inter-chip variability and same-chip reproducibility. The variability ensures that distinct devices produce different outputs given the same input. Reproducibility, on the other hand, is valuable for predictability and determinism in the device authentication behavior. As a result, PUFs based on complex physical systems provide significantly higher physical security over the traditional systems which rely on storing secrets in nonvolatile memory. In addition, special manufacturing processes are not required to produce PUF devices. This advantage makes PUF devices a cost-effective and reliable alternative to mathematical randomness sources. 

\begin{figure} 
\centering
\includegraphics[width=1\columnwidth]{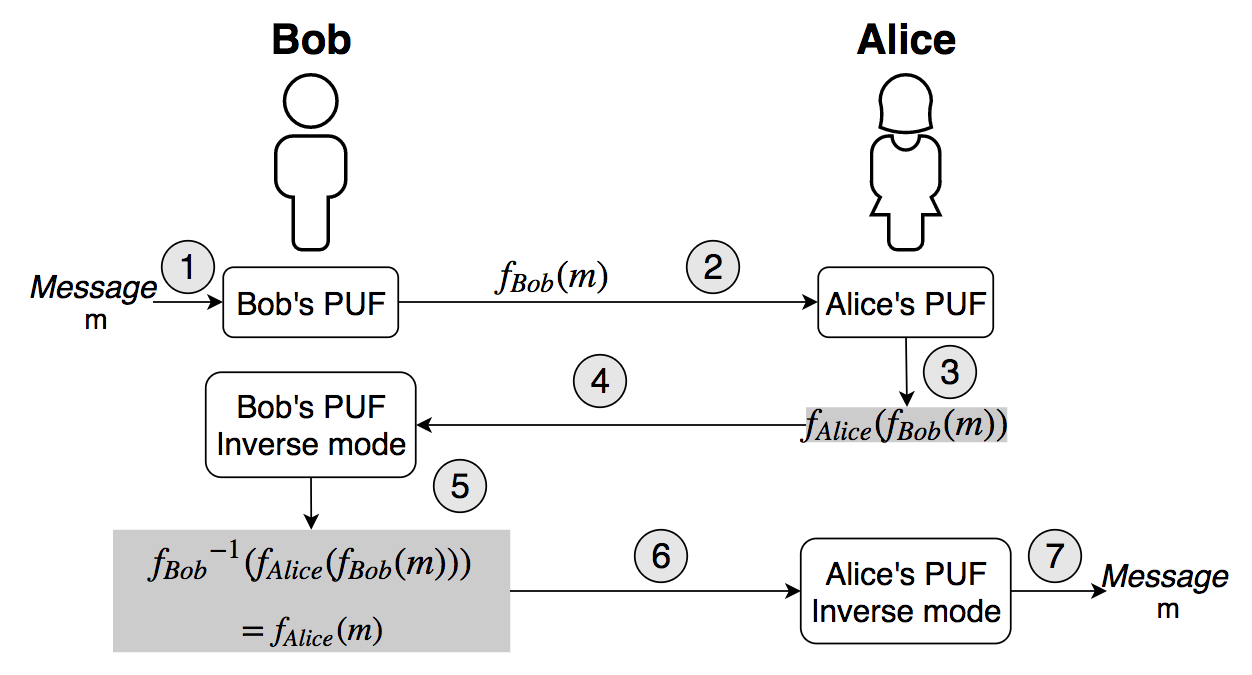}
\caption{\textcolor{red}{Encryption protocol with message encryption based on commutative PUFs $f_{Bob}$ and $f_{Alice}$.}}
\label{fig:protocal}
\end{figure}


So far, the use of PUFs in cryptography is somewhat limited - the most common being key generation or random number generation. Chen used analog circuits to support cryptography with some elements of PUF like randomness \cite{chen2009analog}. Choi \textit{et al.} deployed a variant of arbiter PUF to replace symmetric encryption in RFID domain as an authentication mechanism  \cite{choi2010puf}. This was based on the earlier work of Suh \textit{et al.} that deployed PUFs for anti-counterfeiting in RFIDs \cite{devadas2008design}. Che \textit{et al.} described another authentication protocol based on PUFs \cite{che2015puf}. \cite{cryptoeprint:2016:Chat} developed an IoT communication protocol based on PUFs. \cite{kleber2015secure} developed a code encryption engine based on PUFs for supporting a secure execution environment similar to AEGIS \cite{suh2005design}. The key difference between a processor secure execution environment and general encryption is that for the former scenario the processor platform is 
both the source and destination for the communication. In a processor secure execution environment, both the sender and receiver have access to the same physical PUF on the same platform. However, for general encryption, this assumption is violated. Both the sender and receiver possess distinct and different PUFs. We show a general communication protocol based on commutative PUFs.

The key contributions of this paper are: \textcolor{red}{(1) We explore several PUFs based information exchange protocols which serves to encrypt/decrypt information, find the best protocol through analysis; (2) this protocol requires PUFs to be physically commutative.
We develop a framework for physically commutative PUFs based on permutation networks;
(3) We evaluate permutation networks based physically commutative PUF framework with
a primitive permutation network using barrel shifters. Barrel shifters have symmetric input to output
path delays. Hence if two different paths within the same barrel shifter generate randomly uncorrelated 
delays, it is a strong lower bound for randomness in general permutation networks with more skewed
path delays; (4) The results show good same chip, same path delay reproducibility; good differentiation between different chip, same path delay and same chip, different path delay; delays within 1-bit accuracy 
for the logic high and logic low propagation through the same path demonstrates physical commutativity; and 
good pseudo-random number generator properties for delay.}

 This paper is organized as follows. Section~\ref{sec:protocol} introduces communication protocol. Section~\ref{sec:pufschema} illustrates commutative PUF encryption protocol. Section ~\ref{sec:bigpicture} shows the schematic of barrel shifter PUF. Section~\ref{sec:circuitdesign} presents the detailed circuit implementation of barrel shifter PUF. Variability/reproducibility and commutativity test results based on post-layout simulations are presented in Section~\ref{sec:simulationresult}. Performance of BS-PUFs based encryption protocol is evaluated in Section~\ref{sec:modeling}. Sections \ref{sec:future} and \ref{sec:conclusion} discuss future work and conclusions.
\section{Communication Protocol}
\label{sec:commuprotocol}
\label{sec:protocol}

\tim{
Fig. \ref{fig:protocal} depicts Bob as the sender and Alice as the receiver. 
Both Bob and Alice have their own PUF. 
If Bob encrypts his message $m$ with his PUF as $f_{Bob}(m)$, Alice has no way to decrypt it except
to ask Bob to decrypt it for her. 
The following protocol overcomes this asymmetry.
}


\begin{enumerate}
\item  Bob encrypts the message $m$ with $f_{Bob}$.
\item  Bob sends $f_{Bob}(m)$ to Alice.
\item  Alice encrypts $f_{Bob}(m)$ with $f_{Alice}$. (At this point, Alice does not know the message $m$.)
\item  Alice sends $f_{Alice}(f_{Bob}(m))$ to Bob.
\item  Bob decrypts $f_{Alice}(f_{Bob}(m))$ with $f^{-1}_{Bob}$ and obtains $f_{Alice}(m)$.
\item  Bob sends $f_{Alice}(m)$ to Alice.
\item  Alice decrypts $f_{Alice}(m)$ with $f^{-1}_{Alice}$ and obtains the message $m$.
\end{enumerate}

\tim{
Message confidentiality is maintained by entangling message bits with physical randomness.
The entangling process must be \textit{commutative} so that the order of $f_{Alice}$ and $f_{Bob}$ can be changed.
Decryption of entangled messages requires \textit{reversibility}.
The entangled message $m'$ must exhibit a \textit{non-linear relationship} with $m$;
this makes it hard for an eavesdropper to learn $m$ by examining intermediate messages.
}



\begin{figure} 
\centering
\includegraphics[width=2.5 in]{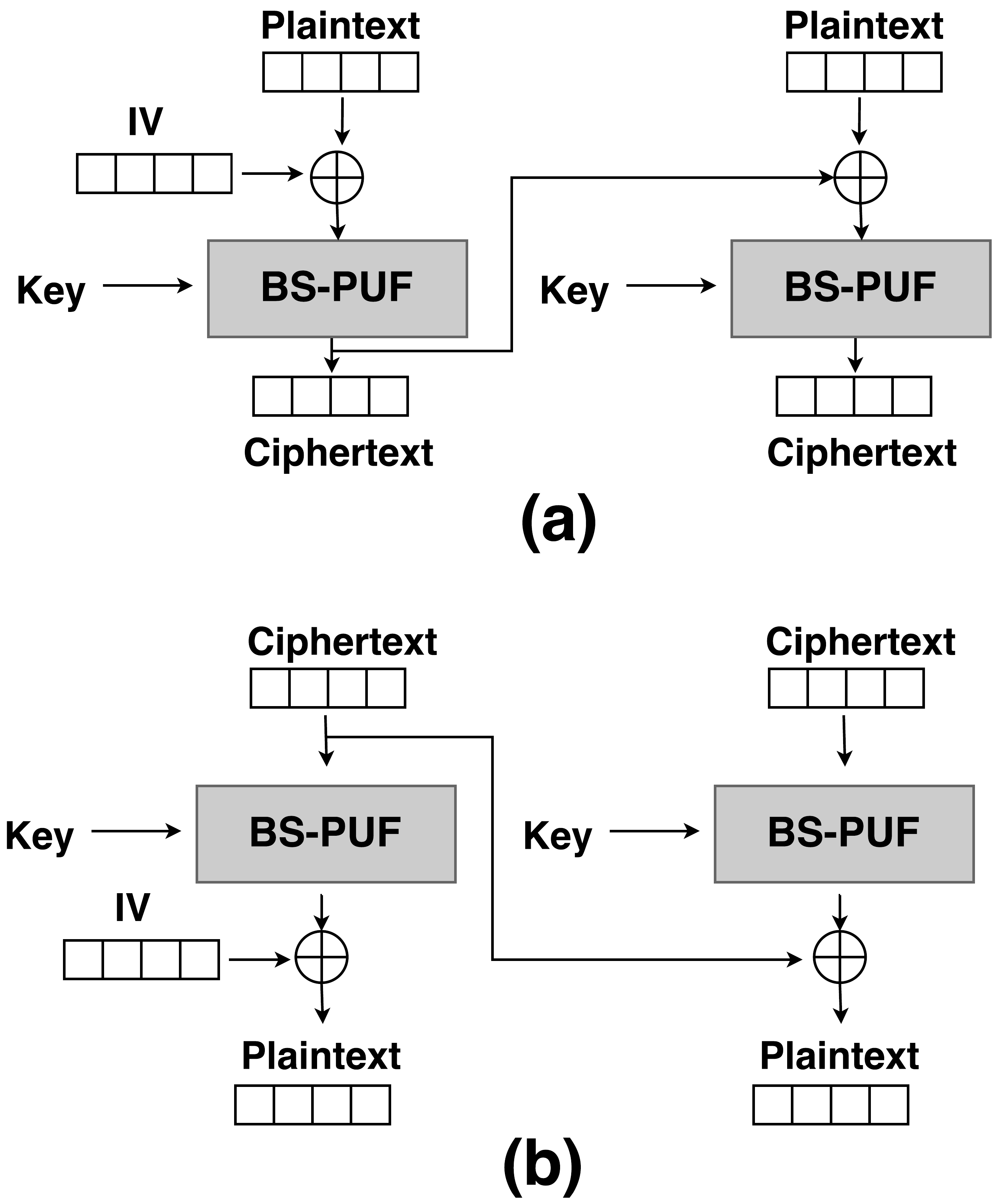}
\caption{
\tim{
Cipher block chaining methods are used to encrypt (a) and decrypt (b) messages.
This prevents the adversary from identifying plaintext patterns;
it ensures identical blocks of plaintext encrypt to different ciphertext.
}
}
\label{fig:encryptprotocol}
\end{figure}

\tim{
The circuit design and encryption protocol enable the commutative, invertible, and non-linear relationship properties of messages.
Section~\ref{sec:pufschema} describes a mechanism for BS-PUF-based encryption.
The BS circuit design is detailed in Sections~\ref{sec:bigpicture}, \ref{sec:circuitdesign}.
}


\section{Encryption Protocol}
\label{sec:pufschema}

\tim{
Encryption must entangle the physical randomness of BS-PUF with the message.
Physical randomness is extracted by measuring the delay of message bits along a shift path.
An XOR of the message bits and delay accomplishes entanglement;
this allows for commutativity and reversibility.
}


\subsection{Encrypting Large Messages} 


\tim{
A BS-PUF uses an $n$-bit key as shift amount. This allows for a 
a $2^n$-bit BS-PUF challenge (message) resulting in a $2^n$-bit BS-PUF response.
Alternately, one could view $(n-bit \; key, 2^n-bit \; message)$ as a challenge. 
We take the former $2^n$-bit challenge view in this paper. For a barrel-shifter,
practical values for $n$ are limited to be in the range $7-10$ bits leading to 
a message block size of $128-1024$ bits. This means that
a method of entanglement/encryption for plaintexts greater than $2^n$ bits is needed.

Entanglement could occur by serializing the blocks of plaintext at BS-PUF input and concatenating the generated ciphertexts.
However, this approach reveals patterns in the plaintext;
the same plaintext will always encrypt to the same ciphertext.
This leaks information by allowing an adversary to identify plaintext patterns.

The technique of cipher block chaining (CBC) is typically applied in block ciphers such as AES \cite{daemen2013design}.
Like AES, BS-PUF encrypts a fixed number of plaintext bits.
Thus, it can be viewed as a block cipher. A practical barrel-shifter or permutation network implementation could 
consist of 128-1024 bit blocks.


Fig. \ref{fig:encryptprotocol} applies CBC to two blocks of plaintext.
Before applying BS-PUF, the plaintext $p_i$ is XOR'ed with the previous ciphertext $c_{i-1}$.
The output of BS-PUF using key $K$, $BS-PUF(p_i, K)$, is the ciphertext, $c_i'$.
Thus, encryption of the $i^{th}$ block is $c_i = BS-PUF(p_i \oplus c_{i-1}, K)$.
The result is a cipher text $c_1 || c_2 || \dots || c_m$ for $m$ blocks where $||$ denotes concatenation.

$c_0$ is an initialization vector (IV).
This IV must be updated with each message;
otherwise the same plaintext will encrypt to the same ciphertext.
This would again allow an eavesdropper to identify patterns. \textcolor{red}{Unlike traditional CBC algorithms, IV for BS-PUFs based encryption does not need to be public because ciphertext will be sent back to sender for decryption. It could be generated with any PUF, e.g. SRAM PUFs \cite{holcomb2009power}.}

Decryption utilizes BS-PUF's inverse.
$p_i$ is recovered by the reverse process.
Ciphertext $c_i$ is given to the inverse BS-PUF operation.
The $\oplus$ of the output and $c_{i-1}$ is then taken.
Thus, decryption of the $i^{th}$ block is $p_i = BS-PUF^{-1}(c_i, K) \oplus c_{i-1}$.




Message encryption requires a secret key.
The key determines the bit shift path; it is used as shift amount.
The BS-PUF response depends both on the challenge (plaintext) and the key.
The key does not change as frequently as the plaintext does.

Some of the desirable characteristics of BS-PUF are as follows.
BS-PUF is fast.
Encryption takes multiple rounds with a traditional block cipher.
BS-PUF makes only one pass through the shifter or permutation hardware.

}

\begin{figure} 
\centering
\includegraphics[width=3 in]{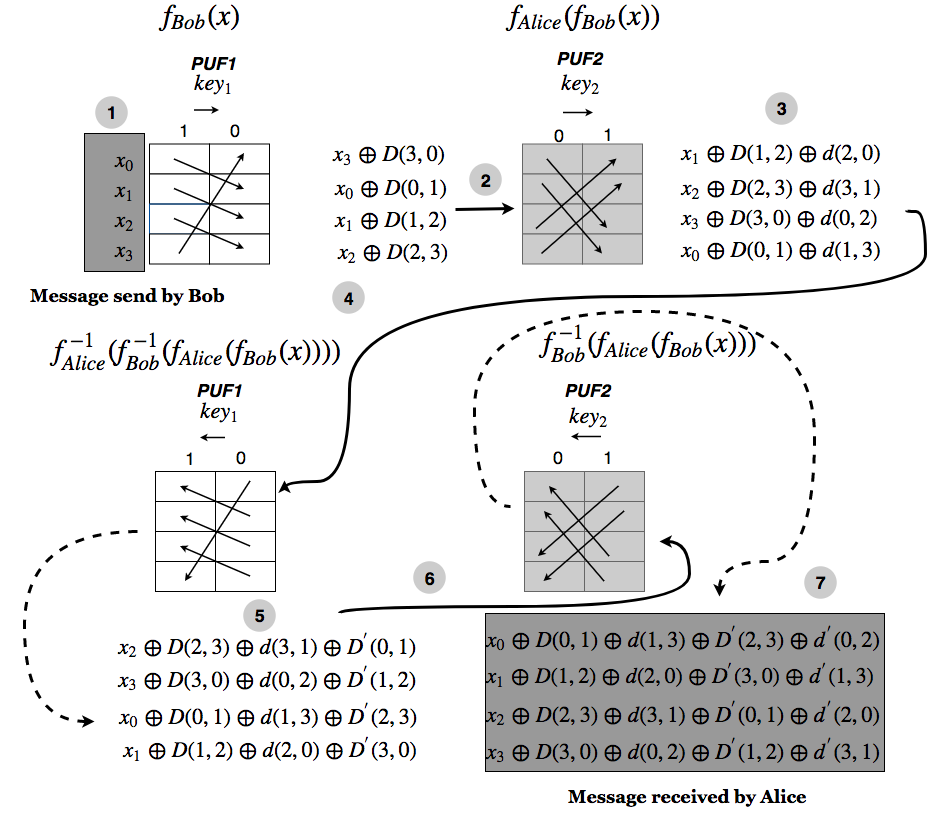}
\caption{
\tim{
(1) Bob applies $f_{Bob}$ and 
(2) sends the result to Alice.
(3) Alice applies $f_{Alice}$ and 
(4) sends the result to Bob.
(5) Bob applies $f_{Bob}^{-1}$ and
(6) returns the result to Alice.
(7) Alice applies $f_{Alice}^{-1}$ hoping to recover the message.
Unfortunately, $f^{-1}$ does not subtract delay from the correct bit in (5), (7);
the correct message is not received by Alice.
This scheme fails to be commutative. 
}
}
\label{fig:wrongschema}
\end{figure}

\subsection{Single Block Encryption}
\label{sec:blockencrypt}
In this subsection, several permutation schemes are discussed for single block encryption.  

\subsubsection{\textcolor{red}{\textbf{Asymmetric Key Encryption}}}
\tim{
Encrypting without a shared key is ideal.

}
\tim{
Section~\ref{sec:protocol} dictates invertibility and commutativity as communication protocol requirements.
}


PUF $f$ must be a one-to-one function to achieve encryption and invertibility for decryption.
Many classical PUFs, such as RO-PUFs \cite{mansouri2012ring,yin2010lisa,maiti2009improving,maiti2011improved} and arbiter PUFs \cite{hori2010quantitative,tajik2014physical}, 
cluster the challenges into equivalence classes on a set of attributes resulting in the same response per challenge
equivalence class.
Arbiter PUF uses relative bit arrival time as the clustering attribute.
RO PUF uses relative oscillator frequencies. The end result is that this makes these PUFs not invertible, since the
mapping is many-to-one.

Further note that physical invertibility is distinct from logical invertibility. A mathematical one-to-one function
has logical invertibility, but may not be physically invertible. Physical invertibility is applicable to the PUF
physical attribute measurement process. In the forward computation, inputs traverse the computation paths to the output;
physical measurements may take place at various points along these paths. In the inverse computation, output bits
travel to the inputs through the identical computation paths in reverse. The physical measurements of the same physical
attribute occur in the inverse computation. These forward and inverse physical measurements need to be reproducible
at all measurement points from input to output.

\begin{figure} 
\centering
\includegraphics[width=3.5 in]{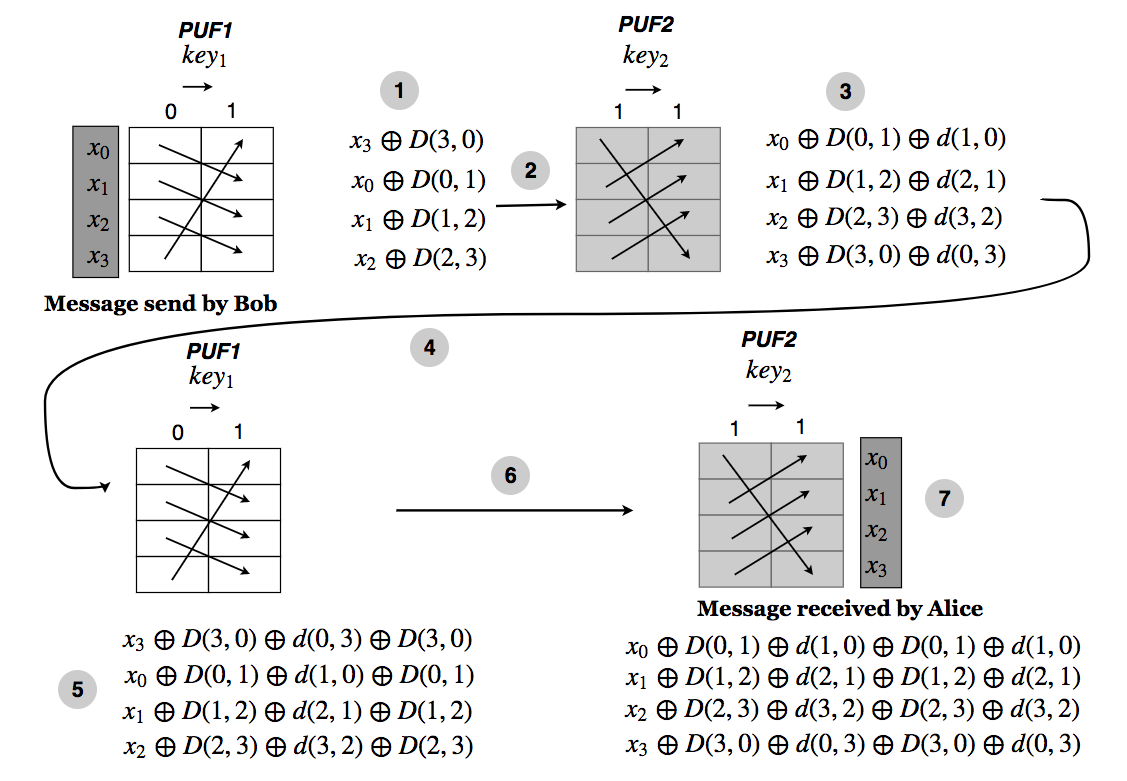}
\caption{
\tim{
Sharing a key allows both parties to perform the same permutation.
This ensures the delay is subtracted from the correct bit when performing the inverse $f_{PUF_l}^{-1}$ for $l=1,2$.
}
\textcolor{red}{
Entropy is added into public message by bit shifting.}}
\label{fig:sharekey}
\end{figure}


\begin{figure*} 
\centering
\includegraphics[width=7in]{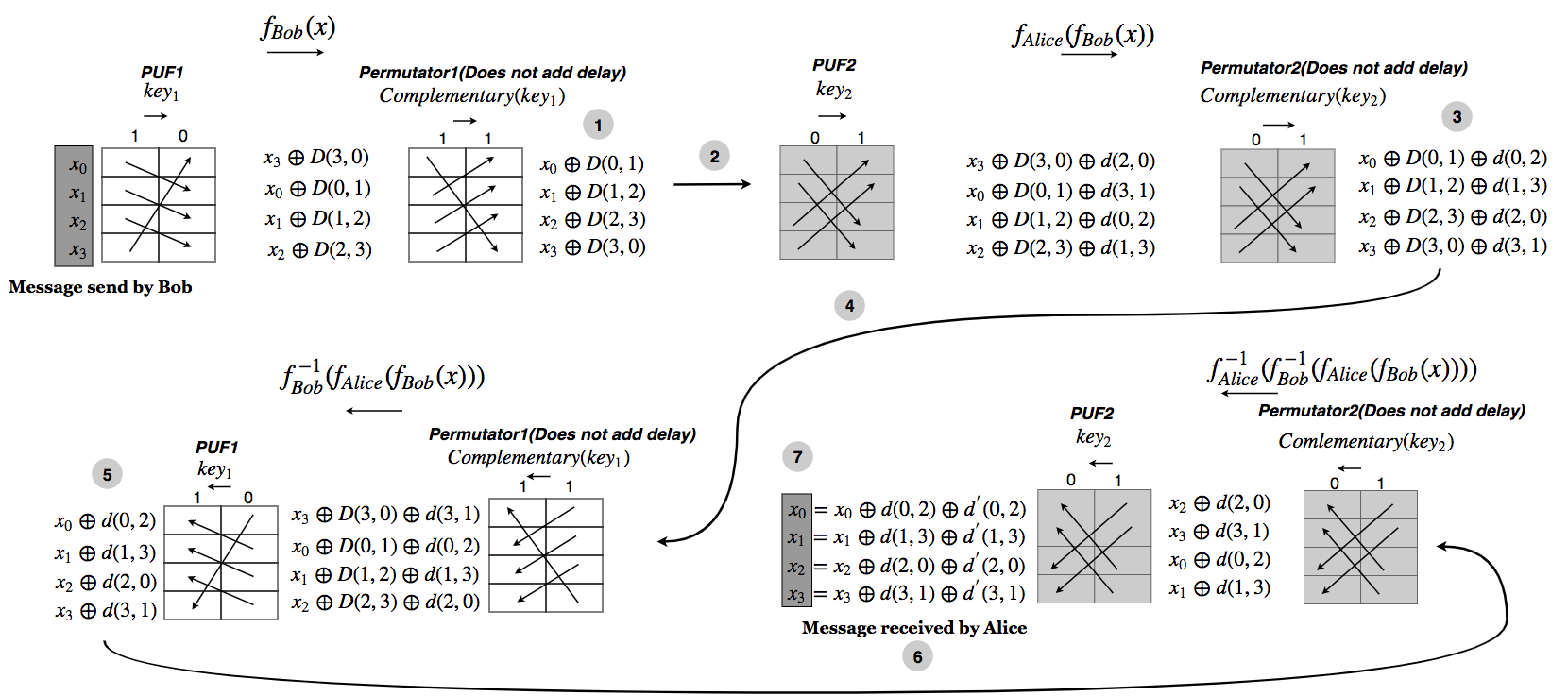}
\caption{
Invertible and Commutative PUF protocol:
\textcolor{red}{$PUF_1$($f_{Bob}$) and $PUF_2$($f_{Alice}$) illustrate the PUF composition and how barrel shifter PUF is used for encryption and decryption processes. Assume both $PUF_1$ and $PUF_2$ are two stages BS-PUFs, $key_1$($PUF_1$)  is $(1,0)$, $key_2$($PUF_2$) is $(0,1)$. For $PUF_1$, bit $x_0$ ($x_1$) goes to output bit position $y_1$ ($y_2$). The encrypted bit output at $y_1$ ($y_2$) is $x_0 \oplus D(0,1)_{m}$ ($x_1 \oplus D(1,2)_{m}$). $D(i,i^{'})_{m}$ is the $m$th least significant bit of the delay from input bit $i$ to the output bit $i^{'}$. Permutator is added after each PUF to shift each bit back to its original position after encryption.}}
\label{fig:pufschema}
\end{figure*}

\tim{
Permutation functions provide the necessary one-to-one relationship.
Permutations create a non-linear relationship from input bits to output bits.
Due to this property, an adversary cannot create a useful mathematical model describing the input, output relationship.
For a $n$-bit data, there exist $N = n!$ permutations denoted by $\pi_0,\pi_1,...\pi_{N-1}$. 
Each $\pi_i$ captures some permutation $( i_0, i_1, \dots, i_{n-1} )$, where bit $k \mapsto i_k$. 
In other words, the bit at $0$ is routed to bit position $i_0$ in the output. 
A key $K$ is used to select this mapping.
We call this a keyed PUF: $R_{i,K} = f(K,P_i)$.
The PUF response is derived from the shift path delay.

The protocol requires the entanglement procedure to be commutative.
Entanglement adds a bit from the delay of each path to the plaintext.
Thus, entanglement is expressed as $f(K_{Bob},P_i) = P_i \oplus D_{Bob}$.
This is commutative because '$\oplus$' is commutative. Note that the entanglement between the physical delay
attribute and logical bits can occur at multiple points during the flight of message bits from input to output;
each measurement point is also an entanglement point.
}

\textcolor{red}{Our first version of encryption protocol is based on invertible and commutative PUFs.}
\tim{
%
Invertibility requires using a raw physical property like delay.
The reversible computation principle states that any information loss makes a process irreversible \cite{bennett1985fundamental}.
Many PUFs derive their response through the comparison of physical properties.
Arbiter PUF uses a race between two paths.
RO-PUF uses a frequency comparison.
These comparisons provide reproducibility by including a wide margin of noise before 
comparison output changes, but information is lost.
}


\tim{
The proposed PUF is based on a barrel shifter.
Constructing it with precisely sized transmission gates makes its delay independent of bit state 0 or 1.
Bit propagation delay for forward path and inverse path is remarkably stable and consistent regardless
of bit state.
This is due to symmetric physical structure of MOSFET's source and drain. 
As we discuss in the following, physical commutativity and invertibility in our protocol is only achieved if the physical
delay on the paths is bit state independent. The Step 5 of Fig. \ref{fig:protocal} when Bob computes
$f_{Bob}^{-1}$ is dealing with a different bit pattern at the output of Bob's PUF than what was computed
in Step 1 at Bob's PUF's output. This is because the Step 5 bit pattern has an additional permutation
applied to it by Alice, which is not known to Bob.
An alternative implementation could have used pass transistors.
However, it is hard to equalize the delay for $0$ and $1$ through a pass transistor.
Thus, transmission gates are used to make the delay plaintext independent.
}

%



Asymmetric key encryption protocol in Section~\ref{sec:protocol} is based on invertible and commutative BS-PUFs;
which are defined as follows:

    \noindent{\bf Invertible PUF:} An invertible keyed PUF $f$ on input $x$ and key $K$: for $f(x, K) = y$ $\implies$
		$f^{-1}(y, K) = x$, where $f^{-1}$ is computed on the same PUF in the reverse direction. Note that the PUF function $f$ entangles a logical component and a physical component,
		and both need to be invertible. 
		
    PUFs designed to be used directly for encryption need two input sequences:
    (1) key for response function selection as in a permutation selector,
    (2) plaintext to be encrypted. 
		
    \noindent{\bf Commutative PUF:} 
    Assume there is a composition of two commutative PUFs PUF1 and PUF2. 
    This means $PUF2(PUF1(x)) \; = \; PUF1(PUF2(x))$. Note that both logical and physical commutativity are needed
		for such a commutative PUF.
    For BS-PUF, the entanglement function must be commutative for physical commutativity in addition to the
		physical measurements being the same in $PUF2(PUF1(x))$ and $PUF1(PUF2(x))$;
    this requires the physical measurements to be invariant of the bit state. The physical measurements are
		completely defined by the key $K$ for a given PUF.

\subsubsection{Protocol Without Permutation}
\label{sec:original}

In the first version of design, each PUF $f_{PUF_1}$ and $f_{PUF_2}$ is a permutation network keyed by $key_{1}$ and $key_{2}$ respectively. 
Key $key_{1}$ selects a permutation $\pi_{key_{1}}$ from a large set of possible permutations - Keccak permutation \cite{bertoni1}, \cite{bertoni2} could be used for instance. 
The implementation, however, needs to be physically and logically reversible consisting of transmission gates. 
We assume that for a permutation $\pi_{key_1}$ which maps $i$th input bit to the $i^{'}$th output bit and $j$th input bit to $j^{'}$th output bit, we capture the exact delays for each input-output path. 
Let $D(i,i^{'})$ denotes the delay of the path from input $i$ to output $i^{'}$ for $\pi_{key_1}$ in $f_{PUF_1}$. 
Let $D(j,j^{'})$ be defined likewise. 
We will describe how we can capture these delays by using 
timer capture and edge detector functions in Section \ref{sec:circuitdesign}.

\textcolor{red}{
For each PUF, the output bit $y_i$  can be expressed as an entanglement function
 $e(x_{\pi^{-1}_{key}(j)}, D(\pi^{-1}_{key}(j),j))$. 
 Here $e$ is an entanglement function between the bit routed to output $j$ ($x_{\pi^{-1}_{key}(j)}$) and the delay of this path from $\pi^{-1}_{key}(j)$ to $j$. 
 The delay $D(\pi^{-1}_{key}(j),j)$ can be quantized to any resolution of $k$ bits. 
 If we use all of the $k$ bits of $D(\pi^{-1}_{key}(j),j)$ to do encryption at the $j$th output bit, we expand the $n$-bit input to an $nk$-bit output. 
 Assuming we want to retain the same output resolution of $n$-bits, one option would be to perform an XOR ($\oplus$) of the $m$th bit of $D(\pi^{-1}_{key}(j),j)$ with the input bit $x_{\pi^{-1}_{key}(j)}$ to generate $y_{j}$ leading to the entanglement function $y_j = e(x_{\pi^{-1}_{key}(j)}, D(\pi^{-1}_{key}(j),j)_{m})$. 
 XOR is a good choice because it is commutative and associative. 
 Since the least significant bit (LSB) and 2nd LSB of $D(\pi^{-1}_{key}(j),j)$ is likely least correlated with the delay of other paths, we have used them in entanglement. 
 The corresponding simulation results are shown in Section~\ref{sec:simulationresult}.
 }

\textcolor{red}{
Let us assume that the delays of the permutation function $\pi_{key_1}$ in $f_{PUF_1}$ are denoted by $D(\pi^{-1}_{key_1}(j), j)$ for a path from input $\pi^{-1}_{key_1}(j)$ to output $j$ and the delays of the permutation function $\pi_{key_2}$ in $f_{PUF_2}$ are denoted by $d(j,\pi_{key_2}(j))$ for a path from input $j$ to output $\pi_{key_2}(j)$. 
Assume that $\pi^{-1}_{key_1}(j) = i$, $\pi_{key_2}(j) = k$, then the output $z_{k} = (x_i \oplus D(i,j)_m)\oplus d(j,k)_m$ is generated. 
The $m$th least significant bit of $PUF_2$'s delay captured by the $d$ function is XORed with $f_{PUF_1}$'s output.
}

Clearly, the RHS of expression  $z_{k} = (x_i \oplus D(i,j)_m)\oplus d(j,k)_m$  is commutative due to commutativity of operator $\oplus$ - it does not matter whether $f_{PUF_1}$ is applied first or $f_{PUF_2}$ is applied first. However, this commutativity statement is only correct for a specific bit routing, but incorrect for encrypted data. 

Consider $PUF_1$ with $\pi_{key_1} = (0 \mapsto 1,1 \mapsto 2,2 \mapsto 3,3 \mapsto 0)$ for a 4 bit input $x_0,x_1,x_2,x_3$ and $PUF_2$ with  $\pi_{key_2} = (0 \mapsto 2,1 \mapsto 3, 2 \mapsto 0,3 \mapsto 1)$. Composition of $f_{PUF_1} \circ f_{PUF_2} = (0 \mapsto 1,1 \mapsto 2,2 \mapsto 3,3 \mapsto 0) \circ (0 \mapsto 2,1 \mapsto 3, 2 \mapsto 0,3 \mapsto 1) = (0 \mapsto 3,1 \mapsto 0,2 \mapsto 1,3 \mapsto 2)$. 
By going over \tim{the } communication protocol in Fig. \ref{fig:protocal} step by step, 
\tim{a defect becomes apparent}.
\tim{The} complete verification process is shown in Fig. \ref{fig:wrongschema}. 

In the following analysis, permutations are abbreviated according to output positions for simplicity. 
e.g. $(0 \mapsto 1,1 \mapsto 2,2 \mapsto 3,3 \mapsto 0)$ is abbreviated to $(1, 2, 3, 0)$. \textcolor{red}{Assume $\pi_{PUF_1} = (1, 2, 3, 0)$ and $\pi_{PUF_2} = (2, 3, 0, 1).$}

\begin{itemize}

\item \textcolor{red}{\textbf{Step 1:}  Apply $f_{PUF_1}$ to $(x_0,x_1,x_2,x_3)$ \tim{resulting } in $(1,2,3,0)(x_0,x_1,x_2,x_3)$, which equals $(x_3 \oplus D(3,0)_m, x_0 \oplus D(0,1)_m, x_1 \oplus D(1,2)_m, x_2 \oplus D(2,3)_m)$.}

\item \textcolor{red}{\textbf{Step 3:}  Apply $f_{PUF_2}$ to $f_{PUF_1}$'s output as in $(2,3,0,1)(1,2,3,0)(x_0,x_1,x_2,x_3)$. This equals $(x_1 \oplus D(1,2)_m \oplus d(2,0)_m, x_2 \oplus D(2,3)_m \oplus d(3,1)_m, x_3 \oplus D(3,0)_m \oplus d(0,2)_m, x_0 \oplus D(0,1)_m \oplus d(1,3)_m)$.}

\item \textcolor{red}{\textbf{Step 5:}  Now  invert the output. Apply $f^{-1}_{PUF_1}$ to $(2,3,0,1)(1,2,3,0)(x_0,x_1,x_2,x_3)$. $f^{-1}_{PUF_1}$ results in $(1,2,3,0)^{-1}(2,3,0,1) (1,2,3,0) (x_0,x_1,x_2,x_3)$ which equals $(x_2 \oplus D(2,3)_m \oplus d(3,1)_m \oplus D^{'}(0,1)_m, x_3 \oplus D(3,0)_m \oplus d(0,2)_m \oplus D^{'}(1,2)_m, x_0 \oplus D(0,1)_m \oplus d(1,3)_m \oplus D^{'}(2,3)_m, x_1 \oplus D(1,2)_m \oplus d(2,0)_m \oplus D^{'}(3,0)_m)$. $D^{'}(i,i^{'})$ denotes the backward path delay from output $i^{'}$ to input $i$. 
According to post-layout simulations, $D^{'}(i,i^{'})$ is always equal to $D(i,i^{'})$ in BS-PUFs.}

\item \textcolor{red}{\textbf{Step 7:} Further applying $f^{-1}_{PUF_2}$ as in $(2,3,0,1)^{-1} (1,2,3,0)^{-1}(2,3,0,1)(1,2,3,0)(x_0,x_1,x_2,\\x_3)$ results in $(x_0 \oplus D(0,1)_m \oplus d(1,3)_m \oplus D^{'}(2,3)_m \oplus d^{'}(0,2)_m, x_1 \oplus D(1,2)_m \oplus d(2,0)_m \oplus D^{'}(3,0)_m \oplus d^{'}(1,3)_m, x_2 \oplus D(2,3)_m \oplus d(3,1)_m \oplus D^{'}(0,1)_m \oplus d^{'}(2,0)_m, x_3 \oplus D(3,0)_m \oplus d(0,2)_m \oplus D^{'}(1,2)_m \oplus d^{'}(3,1)_m)$. This 
logical result is correct in routing $x_i$ back to the $i$th bit position, but the physical delay terms are completely mixed up and do not cancel each other.}
\end{itemize}

\begin{figure} 
\centering
\includegraphics[width=3.5 in]{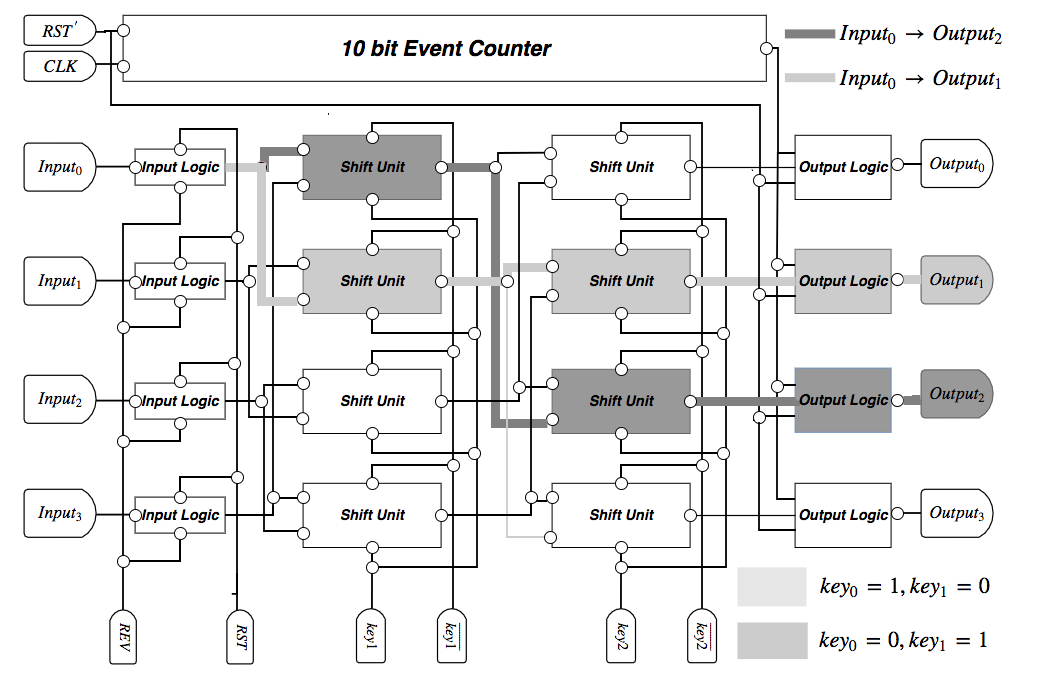}
\caption{\textcolor{red}{
Block diagram of the delay test circuit with two propagation examples. 
When $key_0 = 1$ and $key_1 = 0$, $Input_0$ passes through the light grey path. 
There is one bit shift at the first level and no shift at second level, $Input_0 \to Output_1$. 
When $key_0 = 0$ and $key_1 = 1$, $Input_0$ passes through the dark grey path. 
There is no shift at the first level and there is a two bit shift at the second level, $Input_0 \to Output_2$. }}
\label{fig:schematic}
\end{figure}


\subsubsection{Protocol With Permutation}
\label{sec:modified}
In order to ensure the correct routing and commutativity, we modify the original permutation protocol by adding a permutation after each PUF.
The primary function of this permutation is routing $x_i$ back to the $i$th position from position
$\pi_{key_1}(i)$
before sending the message at the end of Step 1.
The complementary key, $\overline{key_1}$, that results in the
permutation $\pi_{key_1}^{-1}$ is used;
it routes bits back to their original position. Mathematically,
$(\pi_{key_1} \circ (\pi_{\overline{key_1}} = \pi_{key_1}^{-1})) = 1$ where 1 is the identity permutation.
Bit shifting to restore the orginal message bit order is the only function of this permutation.
No delay is added.


An example of this protocol is shown in Fig.~\ref{fig:pufschema} with the following detailed description.

\begin{itemize}
\item \textcolor{red}{\textbf{Step 1:} 
$f_{Bob}$ permutes $x_0,x_1,x_2,x_3$ as in $(1,2,3,0)(x_0,x_1,x_2,x_3)$.
\tim{It computes the physical delay encrypted bit vector, } $(x_3 \oplus D(3,0)_m, x_0 \oplus D(0,1)_m, x_1 \oplus D(1,2)_m, x_2 \oplus D(2,3)_m)$. 
\tim{Before sending it to Alice, Bob's complementary permutation, called permutator in Fig.~\ref{fig:pufschema} is applied to generate }
$(x_0 \oplus D(0,1)_m, x_1 \oplus D(1,2)_m, x_2 \oplus D(2,3)_m, x_3 \oplus D(3,0)_m)$.}

In this new permutation protocol, the logical permutation does not add to the confusion at all unlike in AES or Keccak protocols. 
Confusion is achieved from the permuted physical delay properties of the PUF. 
Which Path delay bits are combined with each input bit is still hidden (through confusion) from the adversary through $key$ driven $\pi$.

\item \textcolor{red}{\textbf{Step 3:} $f_{Alice}$ is applied as $(2,3,0,1)(x_0 \oplus D(0,1)_m, x_1 \oplus D(1,2)_m, x_2 \oplus D(2,3)_m, x_3 \oplus D(3,0)_m)$, resulting in $(x_2 \oplus D(2,3)_m \oplus d(2,0)_m, x_3 \oplus D(3,0)_m \oplus d(3,1)_m, x_0 \oplus D(0,1)_m \oplus d(0,2)_m, x_1 \oplus D(1,2)_m \oplus d(1,3)_m)$. 
\tim{Applying Alice's complementary permutation results in }
$(x_0 \oplus D(0,1)_m \oplus d(0,2)_m, x_1 \oplus D(1,2)_m \oplus d(1,3)_m,x_2 \oplus D(2,3)_m \oplus d(2,0)_m,x_3 \oplus D(3,0)_m \oplus d(3,1)_m)$. }

\item \textcolor{red}{\textbf{Step 5:} Apply $f^{-1}_{Bob}$ to $(x_0 \oplus D(0,1)_m \oplus d(0,2)_m, x_1 \oplus D(1,2)_m \oplus d(1,3)_m,x_2 \oplus D(2,3)_m \oplus d(2,0)_m,x_3 \oplus D(3,0)_m \oplus d(3,1)_m)$.}

\tim{
Decryption follows a similar process.
However, the direction of message transmission is reversed and the inverse permutations are used. This is where
physical invertibility helps recover the original forward delay vector in the reverse direction.
}


\begin{figure} 
\centering
\includegraphics[width=3 in]{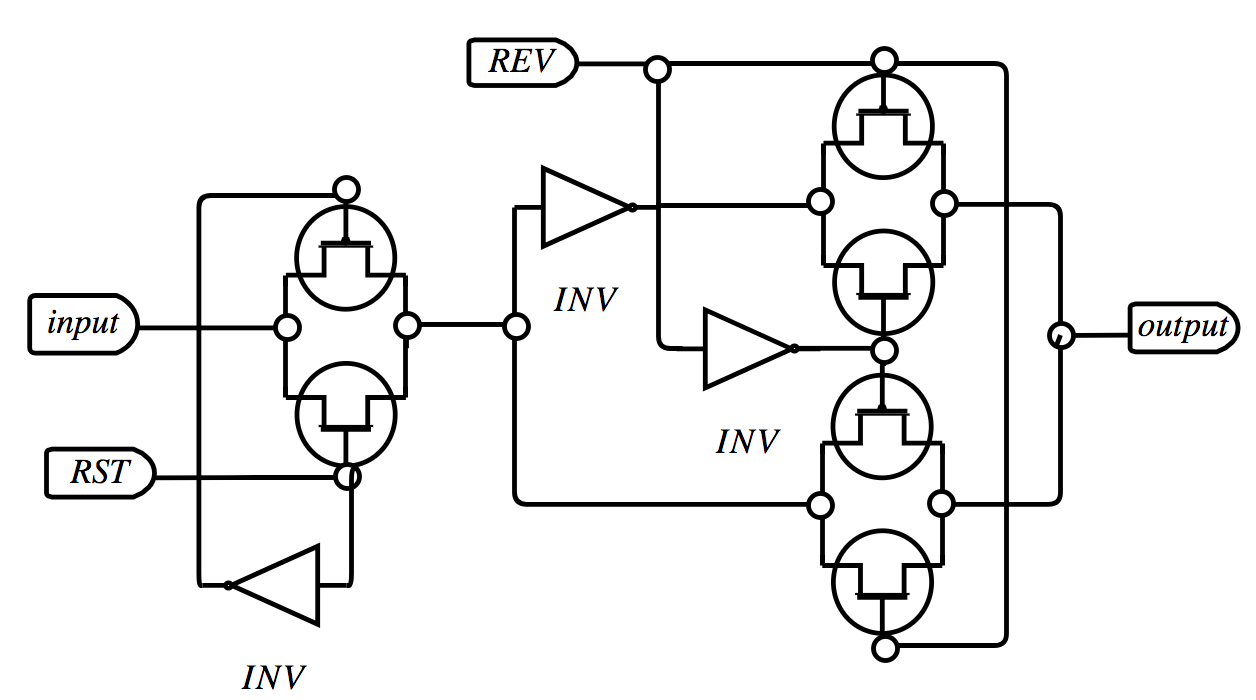}
\caption{\tim{Schematic of 1-bit input logic. Each input bit is controlled by an input logic unit.}}
\label{fig:inlogic}
\end{figure}

\textcolor{red}{Thus, $(1,2,3,0)(2,3,0,1)(x_0,x_1,x_2,x_3))$ is rearranged by Bob's permutator first. 
This is $(x_3 \oplus D(3,0)_m \oplus d(3,1)_m, x_0 \oplus D(0,1)_m \oplus d(0,2)_m, x_1 \oplus D(1,2)_m \oplus d(1,3)_m, x_2 \oplus D(2,3)_m \oplus d(2,0)_m)$.
This rearranged result is given to to $PUF_1$ resulting in $(x_0 \oplus D(0,1)_m \oplus d(0,2)_m \oplus D^{'}(0,1)_m, x_1 \oplus D(1,2)_m \oplus d(1,3)_m \oplus D^{'}(1,2)_m, x_2 \oplus D(2,3)_m \oplus d(2,0)_m \oplus D^{'}(2,3)_m, x_3 \oplus D(3,0)_m \oplus d(3,1)_m \oplus D^{'}(3,0)_m)$.}

\textcolor{red}{Transmission gates show symmetric delays for forward and backward paths;
$D(i,j)$ always equals $D^{'}(i,j)$. 
}
\tim{
Thus, the delay terms cancel.
The result after applying $f^{-1}_{Bob}$ is equal to $(x_0 \oplus d(0,2)_m, x_1 \oplus d(1,3)_m, x_2 \oplus d(2,0)_m, x_3 \oplus d(3,1)_m)$.
}

\item \textcolor{red}{\textbf{Step 7:} $f^{-1}_{Alice}$ is applied. First, Alice's permutator will rotate 
the bits giving $(x_2 \oplus d(2,0)_m, x_3 \oplus d(3,1)_m, x_0 \oplus d(0,2)_m, x_1 \oplus d(1,3)_m)$.
}
\tim{
Rotated bits are then given to $PUF_2$ in the reverse direction resulting in $(x_0 \oplus d(0,2)_m \oplus d^{'}(0,2)_m, x_1 \oplus d(1,3)_m \oplus d^{'}(1,3)_m, x_2 \oplus d(2,0)_m \oplus d^{'}(2,0)_m, x_3 \oplus d(3,1)_m \oplus d^{'}(3,1)_m)$. 
The delay terms cancel.
Alice receives the original message $(x_0, x_1, x_2, x_3)$ sent by Bob. }
\end{itemize}

\subsubsection{\textcolor{red}{\textbf{Symmetric Key Encryption}}}
\tim{
The original protocol in Section~\ref{sec:original} subtracted the delay from the incorrect bit in the inverse permutation.
The protocol shown in Section~\ref{sec:modified} solves the original problem.
However, it contains a fatal flaw;
Using $\oplus$ for entanglement creates a linear relationship between messages in-flight between Bob and Alice.
An eavesdropper can retrieve the original message from the in-flight messages.
}


\begin{figure} 
\centering
\includegraphics[width=2.2 in]{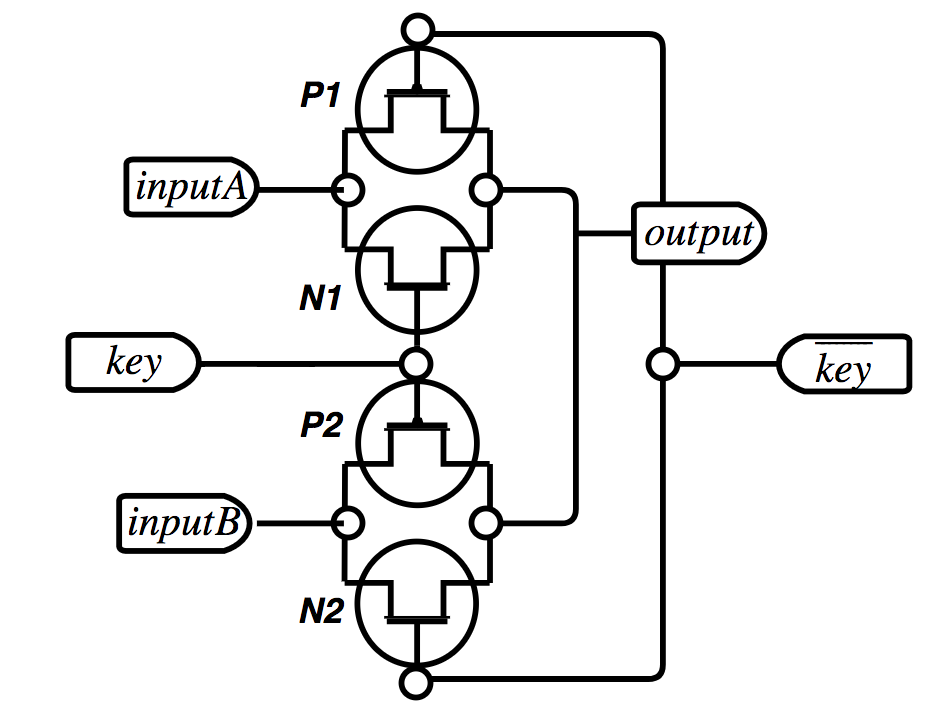}
\caption{Shift Unit of barrel shifter. If $key = 1$ ($\overline{key} = 0$), $N1$/$P1$ is on ($N2$/$P2$ is off), then output equals Input A; otherwise, output equals Input B.}
\label{fig:shiftUnit}
\end{figure}

\textcolor{red}{
Consider Fig. \ref{fig:pufschema} as an example. 
The first bit in original message is $x_0$. 
The encrypted first bit sent from Bob to Alice in Step 2 is $B' = x_0 \oplus D(0,1)$. 
}
\tim{
Then from Alice to Bob in Step 4, $B'' = x_0 \oplus D(0,1) \oplus d(0,2)$. 
The decrypted first bit sent from Bob to Alice in Step 6 is $B''' = x_0 \oplus d(0,2)$. 
$B'$, $B''$ and $B'''$ are all public messages. 
An eavesdropper can extract the original message by: 
}

\begin{enumerate}
\item \textcolor{red}{
Inferring Bob's PUF's delay information by taking XOR of $B''$ and $B'''$. 
$B'' \oplus B''' = x_0 \oplus D(0,1) \oplus d(0,2) \oplus x_0 \oplus d(0,2) = D(0,1)$.
}
\item \textcolor{red}{
Then the original message can be extracted by an XOR of $B'$ and Bob's PUF's delay, $B' \oplus D(0,1) = x_0 \oplus D(0,1) \oplus D(0,1) = x_0$.}
\end{enumerate}

\textcolor{red}{
In order to eliminate this problem, BS-PUF must permute bits in public messages, which we could not do and yet
preserve commutativity and invertibility. 
One possible solution that allows permuted public messages while preserving commutativity and invertibility is 
to let Bob and Alice share the same key.
The corresponding protocol is shown in Fig. \ref{fig:sharekey}. 
}

\tim{
In the shared key protocol, Bob permutes the input message with $\pi_K$ entangling it with his delay.
Alice reverses the permutation using $\pi_K^{-1}$ entangling it with her delay. Note that the shared
key is $K$.
The bits are in their original positions in the message sent to Bob for decryption. 
Note that the entanglement with both PUFs' delays protects this message. 
The delay will be un-entangled from the correct bits in the subsequent decryption steps. The
bit order is different in the message from Bob to Alice versus in the message from Alice to Bob.
This avoids linear leakage of information in XOR based equations on these two messages.
}


Details of the shared key scheme presented in Fig. \ref{fig:sharekey} are as follows.

\begin{itemize}
\item \textcolor{red}{\textbf{Step 1:} Bob permutes $x_0, x_1, x_2, x_3$ with $\pi = (1, 2, 3, 0)$ and gets $(x_3 \oplus D(3,0)_m, x_0 \oplus D(0,1)_m, x_1 \oplus D(1,2)_m, x_2 \oplus D(2,3)_m)$. 
It is sent to Alice without any further bit level routing;
this achieves bit-level confusion of the public message.
}

\item \textbf{Step 3:} $f_{Alice}$ performs the reverse permutation $\pi^{-1}$ of $f_{Bob}$ and simultaneously applies Alice's delay ($\pi^{-1} = (3, 0, 1, 2)$). 
After $f_{Alice}$ is applied, all bits are rotated back to their original position but each bit is encrypted with two physical delay values. 
In this example, after applying $f_{Alice}$ we get $(x_0 \oplus D(0,1)_m \oplus d(1,0)_m, x_1 \oplus D(1,2)_m \oplus d(2,1)_m, x_2 \oplus D(2,3)_m \oplus d(3,2)_m, x_3 \oplus D(3,0)_m \oplus d(0,3)_m)$.


 \begin{figure} 
\centering
\includegraphics[width=3.5in]{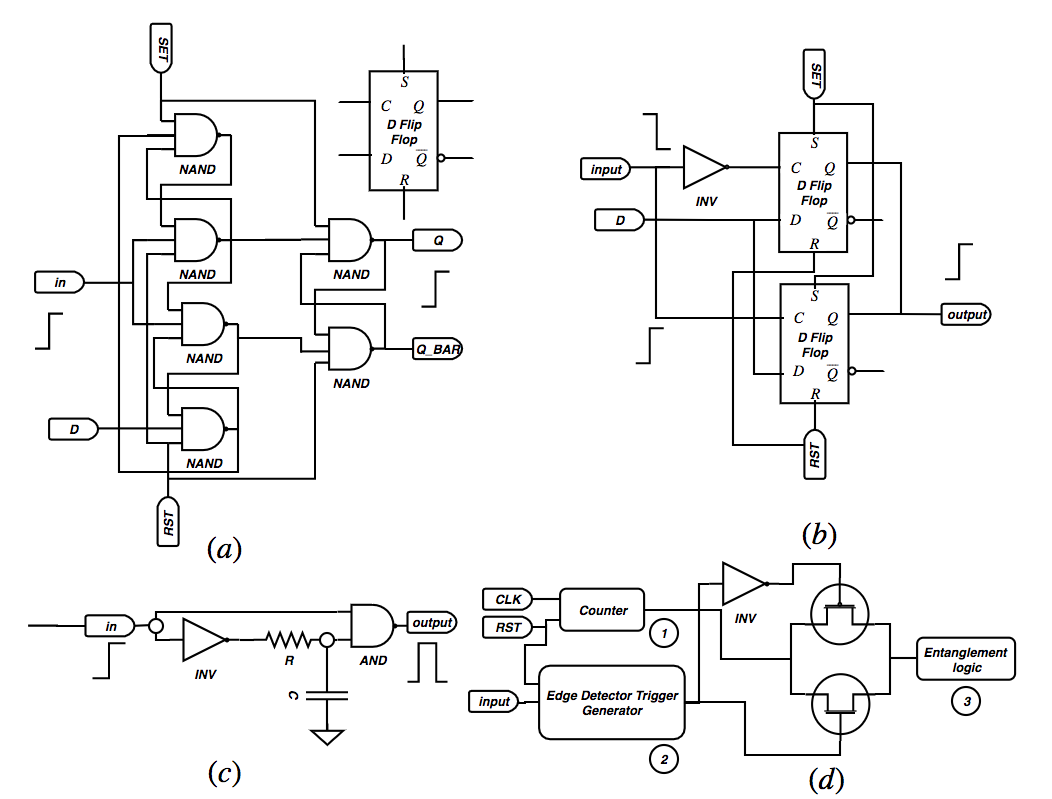}
\caption{
\tim{
{\bf (a) D Flip-Flop} -- 
Triggered by rising edge.
The output, $Q$, is high when there is a rising edge at input, $in$. 
{\bf (b) Edge Detector} -- 
The $output$ reflects a transition at the $input$.
{\bf (c) Positive edge trigger generator} --
Produces a pulse in response to a positive edge at the input, $in$.
{\bf (d) Output Logic} --
Captures the path delay;
this is provided to entanglement logic.
}
}
\label{fig:outlogic}
\end{figure}

\item 
\textcolor{red}{\textbf{Step 5:} $f_{Bob}^{-1}$ is applied. 
Permutation $\pi$ is applied again and delay added in Step 1 is cancelled by XOR. 
Then message sent to Alice is converted to $(x_3 \oplus D(3,0)_m \oplus d(0,3)_m \oplus D(3,0)_m, x_0 \oplus D(0,1)_m \oplus d(1,0)_m \oplus D(0,1)_m, x_1 \oplus D(1,2)_m \oplus d(2,1)_m \oplus D(1,2)_m, x_2 \oplus D(2,3)_m \oplus d(3,2)_m \oplus D(2,3)_m)$ which is $(x_3 \oplus d(0,3)_m, x_0 \oplus d(1,0)_m, x_1 \oplus d(2,1)_m, x_2 \oplus d(3,2)_m)$
}

\item \textcolor{red}{\textbf{Step 7:} $f_{Alice}^{-1}$ is applied, bit positions are rotated back again, and delay added in Step 3 is cancelled by XOR. 
The message from the previous step is converted to $(x_0 \oplus d(1,0)_m \oplus d(1,0)_m, x_1 \oplus d(2,1)_m \oplus d(2,1)_m, x_2 \oplus d(3,2)_m \oplus d(3,2)_m, x_3 \oplus d(0,3)_m \oplus d(0,3)_m)$, which equals the original message $x_0, x_1, x_2, x_3$.}
\end{itemize}

\textcolor{red}{Evaluating all messages crossing the insecure channel, $M' = (x_3 \oplus D(3,0)_m, x_0 \oplus D(0,1)_m, x_1 \oplus D(1,2)_m, x_2 \oplus D(2,3)_m)$, $M'' = (x_0 \oplus D(0,1)_m \oplus d(1,0)_m, x_1 \oplus D(1,2)_m \oplus d(2,1)_m, x_2 \oplus D(2,3)_m \oplus d(3,2)_m, x_3 \oplus D(3,0)_m \oplus d(0,3)_m)$, $M''' = (x_3 \oplus d(0,3)_m, x_0 \oplus d(1,0)_m, x_1 \oplus d(2,1)_m, x_2 \oplus d(3,2)_m)$, no linear relationships exist among any pairs of messages that yield
information to a man-in-the-middle. 
No duplicate delays appear at any bit position. 
There is no way to retrieve original message from the in flight messages without the shared key and access to Bob and Alice's PUFs. }

\tim{
All messages are protected while traversing the insecure channel.
The permutation applied by Bob protects the first message as it travels to Alice.
Entanglement with both Alice and Bob's delay protects Alice's response.
The permutation then protects the final message from Bob to Alice.
}



 \begin{figure} 
\centering
\includegraphics[width=3.5in]{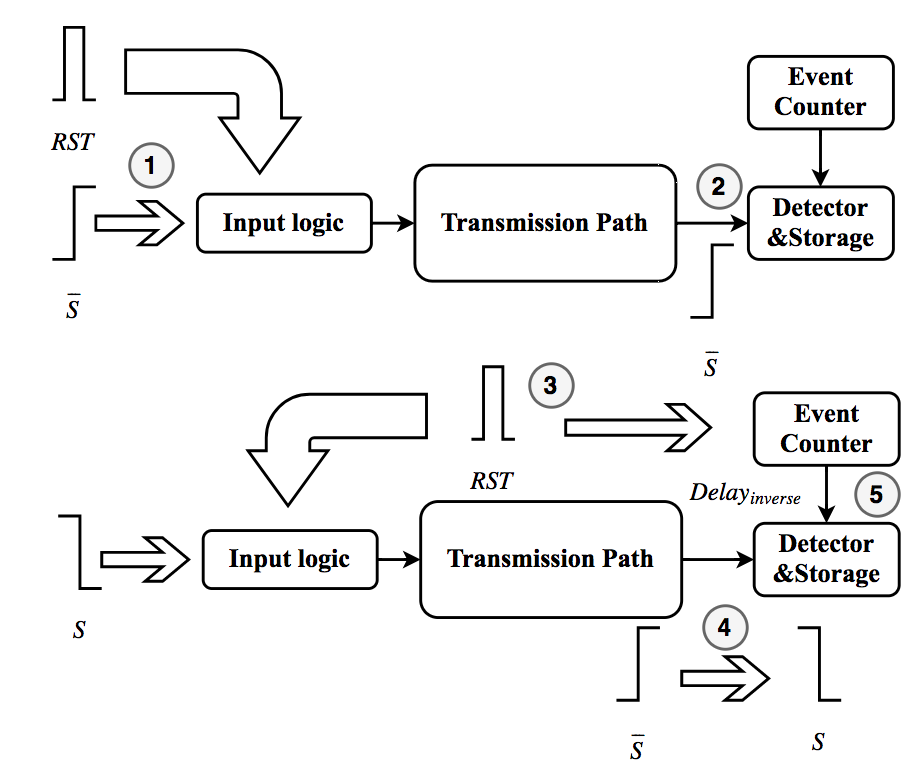}
\caption{
\tim{
The path delay capture unit tests for and stores the path delay.
The edge detector detects an $output$ transition;
$S$ equal to $output$ will not be detected.
Consequently, the transmission path receives $S$ and $\overline{S}$ successively;
a transition at $output$ is guaranteed.
}
}
\label{fig:testprocess}
\end{figure}

\section{Barrel shifter PUF design}
\label{sec:bigpicture}
We evaluate a barrel shifter as a potential invertible and commutative PUF.
The block diagram of a barrel shifter is shown in Fig. \ref{fig:schematic}. 
For simplicity, only two shift levels are shown. 
\tim{
{\it Output Logic} is added to capture path delay $D(i,i')$.
A {\it Event Counter} is initialized to $0$.
The {\it RST} signal simultaneously starts the {\it Event Counter} and releases the input message.
The delay is captured by reading the {\it Event Counter} when the {\it Output Logic} detects a transition.
Finally, the entanglement block in {\it Output Logic} entangles delay information (LSB or 2nd LSB of delay) with the output bit.
}


Each shift stage is logically similar to an arbiter PUF \cite{fruhashi2011arbiter} stage. 

Key bits determine the shift amount $s = \sum_{i=0}^k (key_i * 2^i)$.
Thus, $key_i$ is applied from LSB to MSB, from left to right.
The key determines the shift amount.
For example, in diagram in Fig. \ref{fig:schematic}, 
$key = \{0,1\}$
encodes for right shift by $2$ in the second stage.
Consequently, $Input_0$ traverses a different path;
provides a different delay results with different keys.

The delay variation is generated by transistor-level mismatch \cite{lofstrom2000ic} and doping variability \cite{seoane2009current}. 
Variation accumulates over several stages.
It is then significantly large to be detected by the {\it Output Logic}.


BS-PUF must be invertible;
this property facilitates decryption.
Consequently, the physical delay measurements must not depend on the bit state; they should be
a function only of the path.


\section{Circuit Implementation}
\label{sec:circuitdesign}


\tim{
A commutative PUF based on a barrel shifter is implemented in hardware.
Transmission gates implement the shift paths.
The circuit is subdivided into $3$ components:
}
{\it input logic}, {\it shift unit} and {\it output logic}.

\subsection{Input logic}
\textcolor{red}{{\it Input logic} is used to trigger the delay test system. It is a 3-input, 1-output circuit that connects the input signal $S$ or its inverse $\overline{S}$ to output terminal (Fig. \ref{fig:inlogic}). Input logic consists of three transmission gates. $RST$ (reset) is used to control ON/OFF status of the first transmission gate. When $RST$ is high, $Input$ travels through the first gate and arrives at an intermediate node. Otherwise, it is blocked. $REV$ (reverse) determines whether $Input$ is inverted. $Input$ will be inverted when $REV = 1$. The function definition for input logic is: $output = RST \bullet (REV \oplus input)$. }

\subsection{Shift unit}
\tim{
{\it Shift units} implement the path selection and form shift stages.
Shift unit size determines the magnitude of delay. 
We construct a barrel shifter with $8$ shift stages for testing. 
Each layer $256$ contains shift units. 
Each stage shifts by either $2^{7-n}$ or $0$ where $n$ is the stage index. 
}


\tim{
Each shift unit is a 4-input, 1-output circuit show in Fig. \ref{fig:shiftUnit}.
Either $inputA$ or $inputB$ is mapped to $ouput$.
The mapping is determined by the key.
A $key$ value of $1$ causes the upper transmission gate to open;
$output$ then becomes $inputA$.
Otherwise, $output$ becomes $inputB$. 
}


\begin{table*}[t]
\centering
\caption{NIST TEST RESULTS OF LSB RESPONSE}
{\renewcommand{\arraystretch}{1.2}
\begin{tabular}{|>{\centering\arraybackslash} p{0.4 cm} >{\centering\arraybackslash} p{0.4 cm} >{\centering\arraybackslash} p{0.4 cm}>{\centering\arraybackslash} p{0.4 cm}>{\centering\arraybackslash} p{0.4 cm}>{\centering\arraybackslash} p{0.4 cm} >{\centering\arraybackslash} p{0.4 cm}>{\centering\arraybackslash} p{0.4 cm}>{\centering\arraybackslash} p{0.4 cm} >{\centering\arraybackslash} p{0.4 cm} >{\centering\arraybackslash} p{1.8 cm}>{\centering\arraybackslash} p{1.5 cm} >{\centering\arraybackslash} p{3.5 cm}|}
\hline
c1 & c2 & c3 & c4 & c5 & c6  & c7 & c8 & c9 & c10 & P-VALUE & PROPORTION & STATISTICAL TEST\\
\hline
 16  & 21 & 13 & 19 & 19 & 23 & 16 & 18 & 21 & 34 & 0.099513  &  198/200  &   Frequency\\
\hline
 12 & 24 & 20 & 24 & 12 & 28 & 22 & 15 & 27 & 16 & 0.068999  &  199/200   &  BlockFrequency\\
\hline
 18 & 20 & 20 & 12 & 16 & 19 & 26 & 11 & 31 & 27 & 0.028817  &  199/200  &   CumulativeSums\\
\hline
 17 & 19 & 15 & 14 & 15 & 17 & 34 & 15 & 27 & 27 & 0.011791  &  200/200   &  CumulativeSums\\
\hline
 19 & 14 & 24 & 32 & 16 & 15 & 21 & 18 & 19 & 22 & 0.191687  &  198/200   &  Runs\\
\hline
 19 & 16 & 15 & 16 & 24 & 19 & 25 & 22 & 23 & 21 & 0.769527  &  194/200   &  Serial\\
\hline
 18 &  20 & 20 & 24 & 16 & 17 & 20 & 24 & 24 & 17 & 0.890582  &  197/200  &  Serial\\
\hline
\end{tabular}
}
\label{NISTLSB}
\end{table*}

\begin{table*}[t]
\centering
\caption{NIST TEST RESULTS OF 2nd LSB RESPONSE}
{\renewcommand{\arraystretch}{1.2}
\begin{tabular}{|>{\centering\arraybackslash} p{0.4 cm} >{\centering\arraybackslash} p{0.4 cm} >{\centering\arraybackslash} p{0.4 cm}>{\centering\arraybackslash} p{0.4 cm}>{\centering\arraybackslash} p{0.4 cm}>{\centering\arraybackslash} p{0.4 cm} >{\centering\arraybackslash} p{0.4 cm}>{\centering\arraybackslash} p{0.4 cm}>{\centering\arraybackslash} p{0.4 cm} >{\centering\arraybackslash} p{0.4 cm} >{\centering\arraybackslash} p{1.8 cm}>{\centering\arraybackslash} p{1.5 cm} >{\centering\arraybackslash} p{3.5 cm}|}
\hline
c1 & c2 & c3 & c4 & c5 & c6  & c7 & c8 & c9 & c10 & P-VALUE & PROPORTION & STATISTICAL TEST\\
\hline
 15 & 24 & 22 & 19 & 15 & 17 & 10 & 21 & 20 & 37 & 0.005166  &  200/200  &  Frequency\\
\hline
 12 & 18 & 24 & 27 & 15 & 26 & 20 & 13 & 29 & 16 & 0.048716  &  200/200   &  BlockFrequency\\
\hline
 11 & 21 & 20 & 26 & 16 & 22 & 19 &  9 & 24 & 32 & 0.012650  &  200/200  &   CumulativeSums\\
\hline
 15 & 21 & 15 & 21 & 18 & 18 & 28 & 11 & 28 & 25 & 0.099513  &  200/200  &   CumulativeSums\\
\hline
 22 & 25 & 26 & 20 & 18 & 20 & 16 & 18 & 19 & 16 & 0.807412  &  199/200  &   Runs\\
\hline
17 & 20 & 22 & 21 & 24 & 22 & 18 & 14 & 20 & 22 & 0.917870  &  197/200  &   Serial\\
\hline
24 & 19 & 20 & 19 & 21 & 17 & 18 & 25 & 14 & 23 & 0.825505  &  197/200  &   Serial\\
\hline
\end{tabular}
}
\label{NIST2ndLSB}
\end{table*}

\tim{
The path delay value should vary depending on the shift path.
Path delay primarily depends on shift units' transmission gates.
Adding additional load capacitance after each transmission gate or accumulating variation over several stages of transmission gate enlarge the delay;
it becomes detectable by the path delay counter.

\textcolor{red}{In BS-PUFs, PUFs uniqueness depends on how much delay variation could be provided by same path on different chip.}
Modifying transistor area is the main method for increasing the inter-chip variation.
Transistor delay variation is inversely proportional to transistor area \cite{grunebaum2001mismatch}.
Sizing transistors smaller results in increased delay variation.
However, BS-PUF requires plaintext independent path delay.
Path delay for a $1$-valued bit compared to a $0$-valued bit differs for \textcolor{red}{minimum} transistor sizes.Hence larger transistors are used in shift units.
}




\subsection{Output logic}
{\it Output logic} measures/captures path delay. 
{\it Output logic} for each bit contains $3$ parts: 
{\it counter}, 
{\it edge detector trigger generator} and 
{\it entanglement logic} (Fig. \ref{fig:outlogic}(d)).

\tim{
{\it Counter} takes $CLK$ and $RST$ as input producing a 10-bit output;
it counts the number of rising edges of $CLK$.
Setting $RST$ high resets the counter to $0$.
The path delay is expressed as $($input clock period$) \times ($counter value$)$.
}

{\it Edge detector trigger generator} generates a pulse in response to at transition at its input.
it includes an {\it edge detector} (Fig. \ref{fig:outlogic}(b)) and a {\it positive edge trigger generator} (Fig. \ref{fig:outlogic}(c)). 
\textcolor{red}{{\it Edge detector} converts a rising or falling edge into a rising edge at its output. 
{\it Positive edge trigger generator} converts the rising edge from edge detector into a pulse}.

\tim{
The {\it output logic} works as follows. 
\textcolor{red}{First, a rising/falling edge at input produces a pulse at {\it edge detector trigger generator} output.} 
This pulse enables the transmission gate in Fig. \ref{fig:outlogic}(d) for a short time period ($2ns$). 
During this time, counter output is captured;
it must not change while being captured.
Thus, enable time period must be shorter than clock period ($4ns$).
}
\textcolor{red}{{\it Entanglement logic} extracts the $m$th LSB of delay $D(i,i^{'})$.}
\tim{
Computing XOR of this bit with the input signal $x_i$ results in the entangled output bit.
}

\tim{
The output logic works by detecting a transition.
An transition occurring depends on the previous output value.
Thus, the output logic is incapable of detecting unchanging output values.
An output transition is forced by providing $\overline{x_i}$ before $x_i$ at the input.
}


\tim{
\subsection{Path Delay Testing}
The {\it input logic, shift unit and output logic} work together to capture the path delay.
The following five steps are necessary.
}


\tim{
\begin{enumerate}
\item  Set $\overline{x_i}$ as input and reset {\it input logic}.
\item  Wait for $\overline{x_i}$ to arrive at {\it output logic}. 
\item  Reset {\it input logic} and {\it clock counter}, set $x_i$ as input. 
\item  Wait as $x_i$ travels the path determined by $key$ triggering a transition at the {\it output logic}. 
\item  Encrypt using the captured {\it counter} value.
\end{enumerate}
}


\section{Post-layout simulation result}
\label{sec:simulationresult}
\tim{
The entanglement logic utilizes a $1$-bit result from the path delay.
The path delay capture logic provides a multiple-bit delay counter.
One bit must be chosen;
it must be shown to have the requisite properties for BS-PUF:
(1) inter-chip variability,
(2) intra-chip reproducability,
(3) randomness,
(4) commutativity.

Cadence Spectre simulations are used to generate raw delay data.
Delay variability assessment is conducted by $3\sigma$ Monte Carlo sampling over process parameters. 
This test uses IBM 130 nm PDK. 
A common centroid layout is employed to reduce linear gradient errors \cite{long2005optimal}.
}



\tim{
We construct an $8$-level barrel shifter accepting a $256$-bit input with a $256$-bit output. 
Output logic similar to input capture logic in \cite{microcontroller} detects output voltage changes.
Voltage transitions send a control signal to a counter.
Path delay is captured at the resolution of the counter's clock period; 
a period of $4$ns is used.
Delays must be a reasonable multiple of the clock period to express variation.
}


\tim{
In the following experiments, we primarily focus on raw data: 
(1) Monte Carlo sampling 200 times on the path from input 0 to output 16 
(2) Monte Carlo sampling 200 times on all 256 paths with no shifting.
}

\begin{figure} 
\centering
\includegraphics[width=3.5 in]{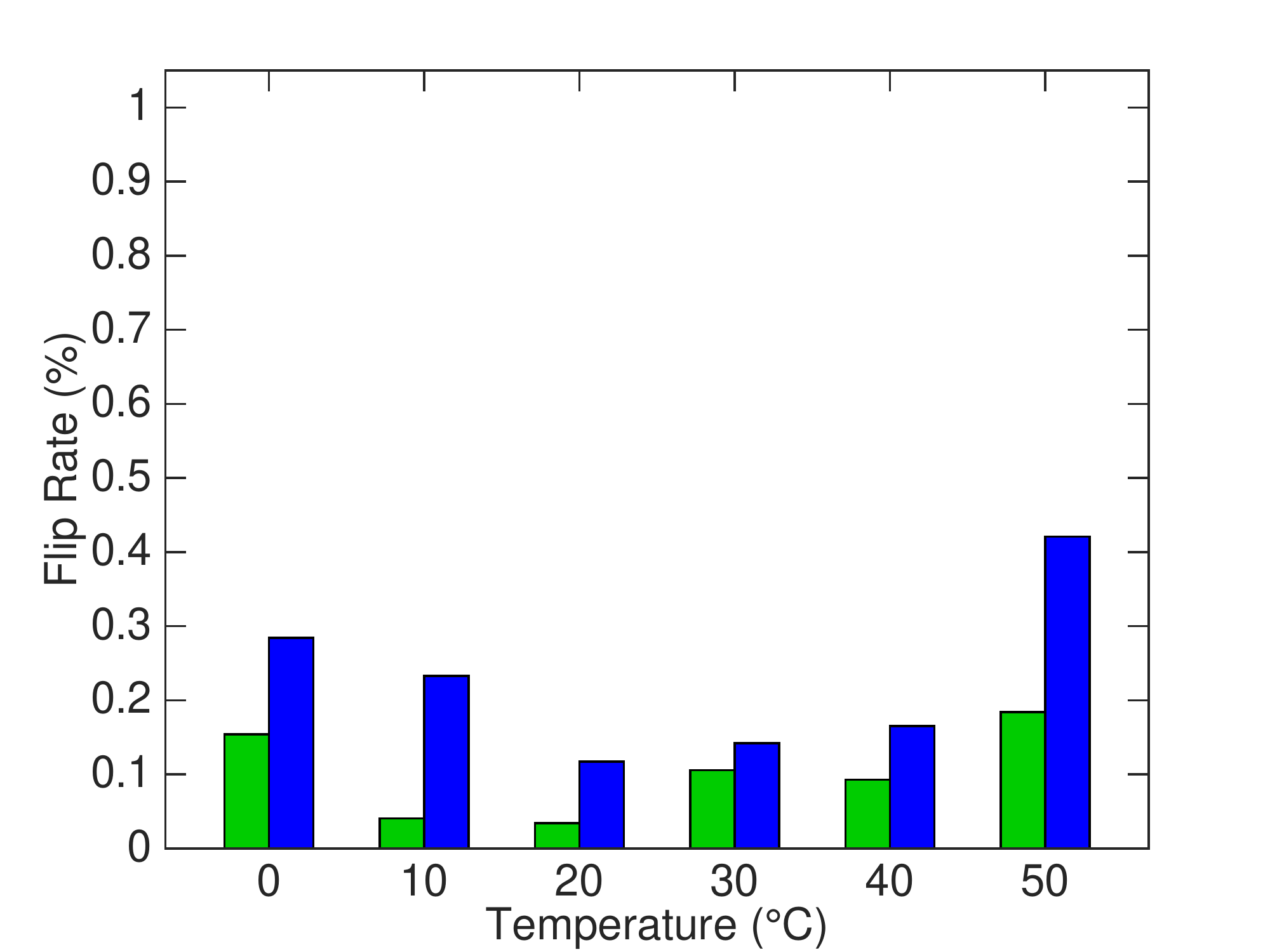}
\caption{\textcolor{red}{Percentage of bit flips under temperature variation. 
Flip rates demonstrate signal-to-noise ratio (SNR) under different temperatures. 
Flip rates of LSB are shown in blue.
Flip rates of 2nd LSB are shown in green. }
\tim{
The flip rate of LSB is much higher than 2nd LSB.
}
}
\label{fig:flipRate}
\end{figure}
\begin{figure} 
\centering
\includegraphics[width=3.5 in]{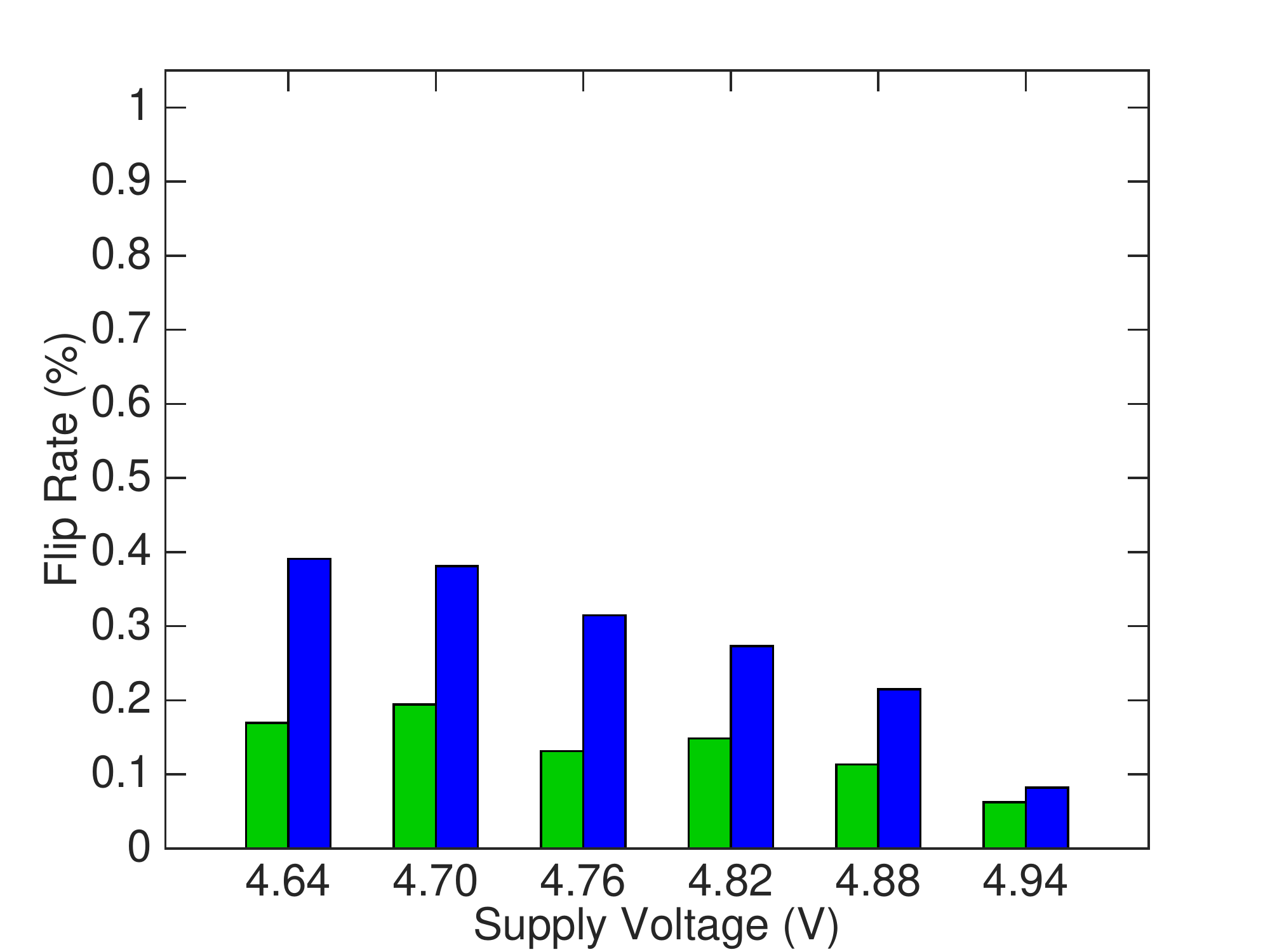}
\caption{\textcolor{red}{Percentage of bit flips under voltage variation. Flip rates of LSB are shown in blue. Flip rates of 2nd LSB are shown in green.}}
\label{fig:voltageflip}
\end{figure}

\tim{
\subsection{Inter-chip Variability}
\label{sec:vari}
Shift path delay is a function of the silicon fabrication process;
it potentially exhibits PUF properties.
Each shift path terminates with entanglement logic requiring one bit.
A bit from the delay counter must be selected.
The chosen bit must exhibit sufficient variation.

Monte Carlo simulation captures single path delay variability as a proxy for inter-chip delay variability. As shown in Fig.~\ref{fig:var}, in $200$ Monte Carlo samples for process parameters perfromed on path $x_0 \mapsto y_{16}$, the delay ranges from $85$ ns to $145$ ns with an average around $120$ ns. 
It is a $\pm25\%$ ($\pm$30-ns) variation. 
Counter output varies about $\pm8$.
This indicates that roughly the least significant $3$ bits of delay have significant entropy in inter-PUF measurements.
Thus, the LSB, 2nd LSB, and 3rd LSB are candidates for entanglement.


\subsection{Inter-chip Uniqueness}
The chosen path delay bit must exhibit inter-chip uniqueness.
This requires significant variance between responses on different chips.
Pair-wise hamming distance (HD) is a criterion that measures variability.


The HD of 200 path delay samples of 256-bit responses is computed.
Table \ref{HDLSB} shows distribution of inter-chip HD for LSB.
Similar figures are given for 2nd LSB in Table \ref{HD2ndLSB}.
\textcolor{red}{
For LSB, the mean HD is $127.99$ bits with a standard deviation of $8.04$ bits. 
For 2nd LSB, these values are $128.01$ bits and $7.99$ bits, respectively.}
HD $128$ means roughly $50\%$ of the response bits differ.
It is maximally unlikely that two BS-PUFs will generate the same output.

\subsection{Intra-chip Reproducibility}
The usefulness of a single PUF relies on it producing a consistent response to a challenge;
they should be independent from the environment.
Tests are performed subjecting BS-PUF to:
(1) temperature variation
(2) voltage supply variation.
The frequency of response bit flips is quantified.

\begin{figure} 
\centering
\includegraphics[width=3.5 in]{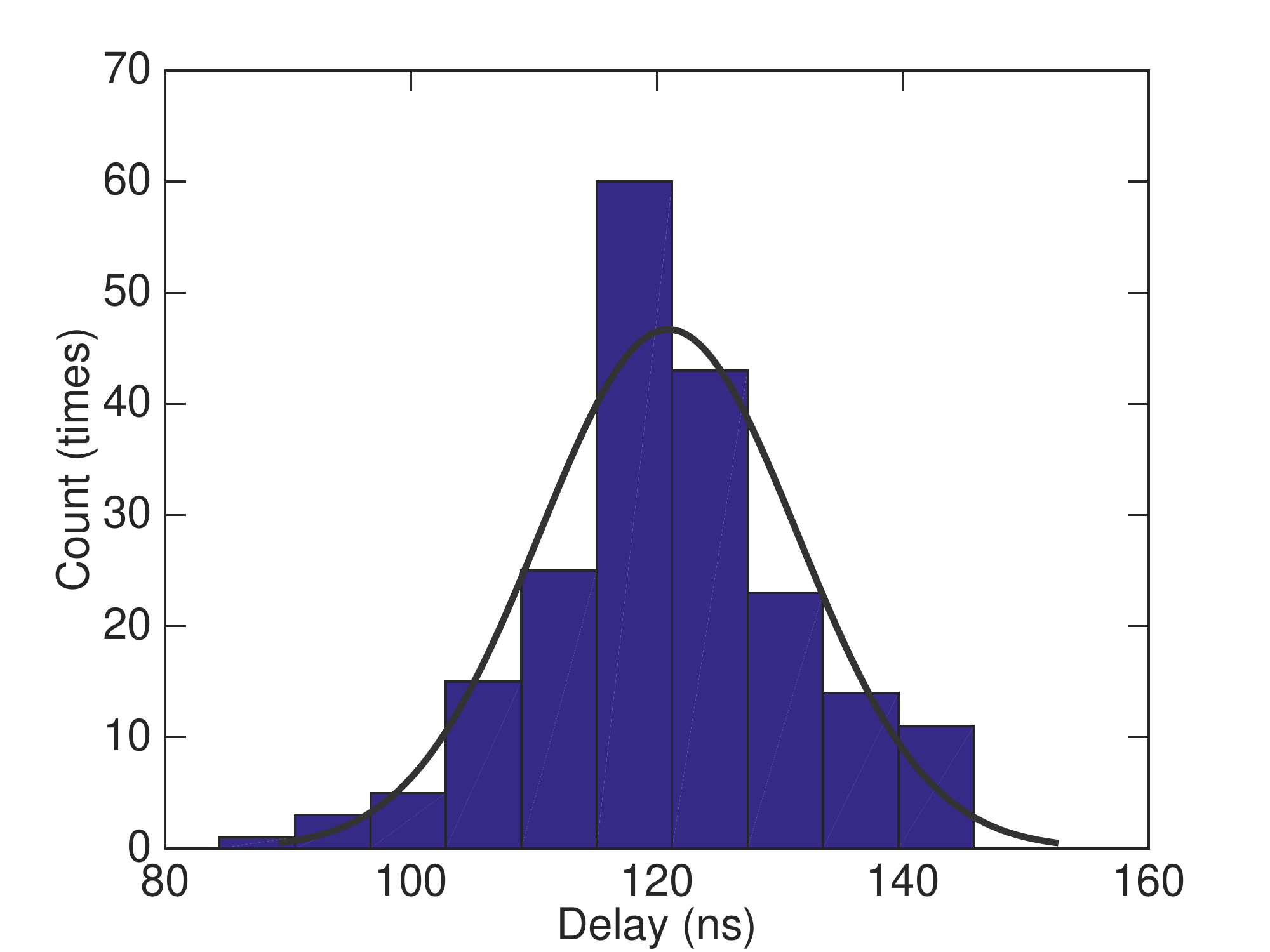}
\caption{Histogram for simulated forward path ($x_0 \mapsto y_{16}$) delay distribution. $25\%$ inter-chip variability is shown.}
\label{fig:var}
\end{figure}


Bit flip rate is frequency a bit changes from $0 \mapsto 1$ or $1 \mapsto 0$.
It is computed relative to some baseline response.
Gathering responses at common room temperature ($25^{\circ}C$) and supply voltage ($5V$) establishes this baseline.
The percentage of path delays where a bit flips is the bit flip rate.
For example, the LSB flipping in $64/256$ paths represents a $25\%$ bit flip rate.

BS-PUF retains a bit flip rate smaller than $18\%$ under environment variation.
This is similar to the flip rate of traditional RO PUFs \cite{gao2014highly}.\


\subsubsection{Temperature Variation}
Temperature is varied from $0$ to $50^{\circ}C$.
Path delay of all $256$ bit paths are gathered with Monte Carlo sampling at $0^{\circ}C$, $10^{\circ}C$, $20^{\circ}C$, $25^{\circ}C$, $30^{\circ}C$, $40^{\circ}C$ and $50^{\circ}C$.
The maximum path delay variation is $-4ns$ to $5ns$.
The counter logic increments at $4ns$ frequency;
a $\pm1$ bit change in path delay is expected.




Knowing how temperature variation affects the chosen entanglement bit is ideal;
bit flip rate quantifies this.
It is computed in response to temperature variation, shown in Fig. \ref{fig:flipRate}.
Vertical bars represent bit flips for LSB (blue) and 2nd LSB (green).
2nd LSB flip rates is under $12\%$ while LSB's flip rate is significantly higher.
Thus, the 2nd LSB provides better reproducibility.

\subsubsection{Voltage Supply Variation}
Supply voltage varies under realistic conditions.
Path delay of all $256$ bit paths are gathered with Monte Carlo sampling at supply voltages of $4.64V$, $4.70V$, $4.76V$, $4.82V$, $4.88V$ and $4.94V$.


Bit flip rate is computed in response to voltage variation, shown in Fig. \ref{fig:voltageflip}.
Flip rates for the 2nd LSB is under $18\%$ while LSB rates are significantly higher.
The 2nd LSB again provides better reproducibility;
it is the best candidate for the entanglement bit.

A higher order bit could be selected.
It would have comparatively better flip rates, but reduced variability.
Many mature techniques exist to compensate for temperature and voltage variation \cite{kumar2012design, vivekraja2011feedback}.
These techniques operate at the flip rates expressed by LSB and 2nd LSB.
Thus, the advantage of choosing a higher order bit is minimal.



\subsection{Randomness}
Output of a good PUF should look like a pseudo-random number generator so that an attacker cannot model it easily. 
Assessing randomness performance of BS-PUF uses data from Monte Carlo sampling of path delays.
Delay values are converted to binary responses by extracting the $m^{th}$ LSB bit from the delay.
Each 256-bit response (one bit from each path) is examined using NIST statistical test suite.


Table \ref{NISTLSB} and Table \ref{NIST2ndLSB} give the detailed test results for LSB and 2nd LSB of the BS-PUFs output.
The minimum pass rate for each statistical test is $193$ for a sample size of $200$ binary sequences according to NIST documentation.
Thus, both LSB and 2nd LSB pass the randomness test;
a proportion greater than $193$ is achieved on all tests.



\subsection{Commutativity}
\label{sec:intest}
Encryption and decryption rely on function composition.
Decrypting a message encrypted by both self and another party is required.
The other party may have changed the bit values ($0$ or $1$).
Thus, Delay variation must be independent of the bit value.
An input of $1$ must have the same path delay as an input of $0$.

BS-PUF path delays depend only on the permutation $key$.
Shift units are sized to achieve balanced pullup and pulldown resistance.
Transmission gate NMOS sizing is ${W_n}/{L_n} = {2}/{3}$ 
PMOS sizing is ${W_p}/{L_p} = {1}/{1}$, where $L_n = L_p$.

Two tests are performed to verify pullup and pulldown variability.
\begin{enumerate}
\item Testing rising/falling edge delay in four different (FF, FS, SF, SS) process corners. 
    Transmission time difference for $0$ and $1$ must be smaller than the counter period ($4 ns$).
\item Performing Monte Carlo sampling of path delay for inputs $0$ and $1$. 
    Delays are recorded for all paths without bit shifting.
    No bit flips should occur in the path delay.
\end{enumerate}

Maximum transmission time difference for $0$ and $1$ is $2.34ns$;
this is much smaller than the $4ns$ clock period.
Consequently, no path delay bits flip in Monte Carlo sampling.

}

\begin{table}[t]\small
\centering
\caption{INTRA-CHIP HD OF BS-PUFs LSB \protect\\
(HD: Hamming distance; \%: percentage of bit-stream pairs with certain HD)}
{\renewcommand{\arraystretch}{1.2}
\begin{tabular}{|>{\centering\arraybackslash} p{1.2 cm}| >{\centering\arraybackslash} p{1.2 cm}| >{\centering\arraybackslash} p{1.2 cm}|>{\centering\arraybackslash} p{1.2 cm}|>{\centering\arraybackslash} p{1.2 cm}|}
\hline
HD & $[90, 100) $ & $[100, 110) $ & $ [110, 120) $ & $[120,130) $ \\
\hline
\% & 0.01\% & 1.11\% & 13.46\% & 42.83\%\\
\hline
HD &  $[130, 140) $ & $[140, 150) $ & $ [150, 160) $ & $[160, 170) $ \\
\hline
\% & 35.03\% & 7.17\% & 0.38\% & 0.01\%  \\
\hline
\end{tabular}
}
\label{HDLSB}
\end{table}

\begin{table}[t]\small
\centering
\caption{INTRA-CHIP HD OF BS-PUFs 2nd LSB \protect\\
(HD: Hamming distance; \%: percentage of bit-stream pairs with certain HD)}
{\renewcommand{\arraystretch}{1.2}
\begin{tabular}{|>{\centering\arraybackslash} p{1.2 cm}| >{\centering\arraybackslash} p{1.2 cm}| >{\centering\arraybackslash} p{1.2 cm}|>{\centering\arraybackslash} p{1.2 cm}|>{\centering\arraybackslash} p{1.2 cm}|}
\hline
HD & $[90, 100) $ & $[100, 110) $ & $ [110, 120) $ & $[120,130) $  \\
\hline
\% & 0.12\% & 2.57\% & 15.68\% & 37.12\%\\
\hline
HD &  $[130, 140) $ & $[140, 150) $ & $ [150, 160) $& \\
\hline
\% & 37.29\% & 6.25\% & 0.97\%& \\
\hline
\end{tabular}
}
\label{HD2ndLSB}
\end{table}

\section{Performance Evaluation}
\label{sec:modeling}

\subsection{Modeling Attack}
\tim{
According to \cite{ruhrmair2010modeling}, all examined Strong PUFs under a given size can be modeled with machine learning with success rates above their stability in silicon. 
Consider the barrel shifter in our communication protocol to be a black box.
Attackers know nothing about the key and physical delay of barrel shifter. 
An attacker should not be able to model the relationship from input bits to the output bits.
Such a model provides an eavesdropper information about the plaintext given a ciphertext.
}

\textcolor{red}{
To investigate the resilience of BS-PUFs against modeling attacks, various ciphertexts are generated with different keys and plaintexts for training and cross-validation.}

\tim{
Logistic Regression (LR) \cite{bishop2006pattern} and Evolution Strategies (ES) \cite{back1996evolutionary, schwefel1993evolution} are commonly used to model PUF output.
ES is specialized to modeling PUFs under noisy conditions \cite{ruhrmair2010modeling};
it does not apply when voltage supply and temperature are certain.
}

\textcolor{red}{
Thus, only LR modeling is performed. 
Since the error rate of machine learning prediction decreases with the size of training set, LR modeling is tested for LSB response and 2nd LSB response with a variety of training sets with different sizes.} 

\tim{

Monte Carlo Sampling \cite{robert2004monte} utilizes randomness to generate $n$ challenge response pairs (CRP).
$n$ random keys, $K = \{K_0,K_1,\dots,K_n\}$  are generated.
Responses, $R$, are generated by entangling plaintext, $P$, using these keys, $R_i = BS-PUF(P, K_i)$.
Note that the response is the shift path delay;
this is dependent on the key only.
Hence, the plaintext need not be modified.
This random CRP sample is assumed to be representative of the distribution of all CRP.


Simulating $BS-PUF(P, K_i)$ requires computationally expensive Cadence Spectre simulations.
An efficient method for computing $R_i$ given $K_i$ is needed.
Thus, we apply Monte Carlo Sampling to create a delay matrix, $D$, modeling the delay of all shift paths.
The delay of each shift unit is recorded.
Path delay is then computed by:
(1) summing the delay of all shift units along a path,
(2) dividing it by $4ns$ capture logic resolution,
(3) extracting LSB or 2nd LSB.
Thus, $D$ enables computations of path delays given $K_i$.
}
\textcolor{red}{For example, Eq. (\ref{eq:delaymatrix}) is a sample delay matrix for $4$ inputs, $2$ stage BS-PUFs. $d_{i,j}$ represents exact delay value of top and bottom transmission gates in $i$th row, $j$th column shift unit.}


\begin{equation}
  D = 
   \begin{bmatrix}
   (d_{0,0,{t}}, d_{0,0,{b}}) & (d_{0,1,{t}}, d_{0,1,{b}})\\
   (d_{1,0,{t}}, d_{1,0,{b}}) & (d_{1,1,{t}}, d_{1,1,{b}})\\
    (d_{2,0,{t}}, d_{2,0,{b}}) & (d_{2,1,{t}}, d_{2,1,{b}})\\
    (d_{3,0,{t}}, d_{3,0,{b}}) & (d_{3,1,{t}}, d_{3,1,{b}})\\
   \end{bmatrix} 
   \label{eq:delaymatrix}
\end{equation}

\tim{
Plaintext-ciphertext pairs (PCP) are computed using $D$.
For the delay matrix in Eq. (\ref{eq:delaymatrix}) using a $key=\{1,0\}$ encoding for right shift in the first stage,
the plaintext $(i_0, i_1, i_2, i_3)$ generates the response in Eq. (\ref{eq:result}).
}
\begin{equation}
R = 
\begin{bmatrix}
i_3 \oplus ((d_{0,0,{b}} + d_{0,1,{t}})/4)_m \\
i_0 \oplus ((d_{1,0,{b}} + d_{1,1,{t}})/4)_m \\
i_1 \oplus ((d_{2,0,{b}} + d_{2,1,{t}})/4)_m \\
i_2 \oplus ((d_{3,0,{b}} + d_{3,1,{t}})/4)_m \\
\end{bmatrix}
\label{eq:result}
\end{equation}
\tim{
This process makes extraction of all possible PCP feasible.
}


\begin{table}[t]\small
\centering
\caption{LR ON LSB WITH 6 AND 8 STAGES BS-PUFs}
{\renewcommand{\arraystretch}{1.2}
\begin{tabular}{|>{\centering\arraybackslash} p{1 cm}| >{\centering\arraybackslash} p{1 cm}| >{\centering\arraybackslash} p{1.2 cm}|>{\centering\arraybackslash} p{1 cm}|>{\centering\arraybackslash} p{1.2 cm}|}
\hline
ML Method & Bit Length & Prediction Rate & PCPs & Training Time  \\
\hline
LR & 64 & \tabincell{c}{17.5\% \\ 28.6\% \\ 58.3\% } & \tabincell{c}{800\\ 8,000 \\ 80,000} & \tabincell{c}{0.0203 sec \\ 0.3580 sec \\ 1.3157 sec} \\
\hline
LR & 256 & \tabincell{c}{9.1\%\\ 18.3\% \\ 25.5\%} & \tabincell{c}{1000\\ 10,000 \\ 100,000} & \tabincell{c}{0.0186 sec\\ 0.3670 sec \\ 2.3212 sec}  \\
\hline
\end{tabular}
}
\label{LRLSB}
\end{table}

\begin{table}[t]\small
\centering
\caption{LR ON 2ND LSB WITH 6 AND 8 STAGES BS-PUFs}
{\renewcommand{\arraystretch}{1.2}
\begin{tabular}{|>{\centering\arraybackslash} p{1 cm}| >{\centering\arraybackslash} p{1 cm}| >{\centering\arraybackslash} p{1.2 cm}|>{\centering\arraybackslash} p{1 cm}|>{\centering\arraybackslash} p{1.2 cm}|}
\hline
ML Method & Bit Length & Prediction Rate & PCPs & Training Time  \\
\hline
LR & 64 & \tabincell{c}{43.2\% \\ 52.6\% \\ 79.5\% } & \tabincell{c}{800\\ 8,000 \\ 80,000} & \tabincell{c}{0.0315 sec \\ 0.1658 sec \\ 1.0104 sec} \\
\hline
LR & 256 & \tabincell{c}{32.4\%\\ 41.0\% \\ 62.8\%} & \tabincell{c}{1000\\ 10,000 \\ 100,000} & \tabincell{c}{0.0157 sec\\ 0.4620 sec \\ 1.6245 sec}  \\
\hline
\end{tabular}
}
\label{LR2ndLSB}
\end{table}

\tim{
For a BS-PUF with an input message length of $256$-bit, there are $2^{256}$ possible input messages.
There are $8$ stages with $2^8$ possible keys.
It is infeasible to generate all $2^{264}$ PCPs.
Linear Regression (LR) is performed with a training set of size $n = \{10, 100, 1000\}$ PCPs per key.
To obtain a representative sample of PCPs, responses are computed with $100$ keys and $10,000$ plaintexts.
PCPs not part of the training set are used for cross-validation.
}
\tim{
Scalability experiments are conducted on a $6$-stage, $64$-bit input BS-PUF;
delay matrix of this BS-PUF is the top left $64 \times 6$ sub-matrix of the $8$-stage delay matrix acquired from Monte Carlo Sampling.
The number of CRPs $N_{CRP}$ that are required to learn a $k$-stage arbiter PUF with error rate $\epsilon$ is $0.5 \times (k + 1) / \epsilon$ \cite{ruhrmair2010modeling}.
Thus, for a $6$ stage BS-PUF, we also scale down $n$ to $8$, $80$, and $800$ PCPs per key.
}


\tim{
Table~\ref{LRLSB} and Table~\ref{LR2ndLSB} show the prediction accuracy of LR on LSB and 2nd LSB.
LR is implemented by an iterative program written in Matlab. 
The regression coefficients' initial values are set to $(0, 0)$ in all LR applications. 
Silicon stability of BS-PUFs is $75\%$.
Thus, all modeling reaching a higher prediction rate should be considered a success. 

LSB provides better result than 2nd LSB. 
LR achieves $79.5\%$ prediction rate for $6$-stage BS-PUF 2nd LSB output.
If 2nd LSB is used as the delay bit, 
then LR can successfully model $6$-stage BS-PUF with sufficient number of PCPs.
On the other hand, with the same modeling process, LSB cannot be modeled even with a large number of training samples.
This is expected as the LSB is inherently more variable.
Consequently, the choice to use LSB or 2nd LSB for the delay bit presents a tradeoff between security and reproducibility;
LSB provides the former while 2nd LSB provides the latter.
}


\subsection{\textcolor{red}{Speed Performance}}

\textcolor{red}{One of the most important advantages of BS-PUFs based encryption is its faster encryption than other traditional symmetric encryption schemes, such as AES. In this section, comparison is made between BS-PUFs encryption and AES.}

    BS-PUF based encryption outperforms conventional AES implementations.
    Some exceptions relying on high-speed crypto processors and architectures exist \cite{zhang2004high}.
    Performing AES Encryption on a modern Intel Pentium Pro processor requires $18$ clock cycles per byte. Decryption
		takes even more cycles with a conservative estimate of $36$ clock cycles per byte for encryption/decryption round-trip.
    This time increases as the block size increases. 
    Comparatively, BS-PUF-based encryption ($1.6$ clock cycles per byte per BS-PUF resulting in $6.4$ clock cycles for both encryption/decryption) is an order of magnitude improvement. In addition, BS-PUF-based encryption scales better,
    because encryption delay is near-constant ($\log n$ delay for block size $n$) regardless of block length.
    
    %


\tim{
This work proposes a protocol for data transmission using BS-PUF.
It necessitates multiple-message round transaction between sender and receiver.
This 
incurs transmission overhead.

The BS-PUF protocol has advantages  over AES in encryption speed.

}


\subsection{\textcolor{red}{Area Needs}}


\tim{
BS-PUF does very little mathematical computation;
protection is provided by the physical properties of the encryption device.
Little area is required due to this simplicity.
In comparison, AES performs many more computations requiring greater area.
}

\textcolor{red}{
According to \cite{hamalainen2006design}, 32-bit FPGA-based AES encryption contains $8,300$ 2-input NAND gate equivalents. 
}
\tim{
A $32$-bit BS-PUF requires $2,400$ transistors, which is 
$600$ 2-input NAND gate equivalents.
This evaluation is not technology dependent.
}


\section{Future Work}
\label{sec:future}

Much needs to be addressed to establish the practicality of commutative PUFs. 
An evaluation of PUFs based on more relevant permutation families such as Keccak sponge family \cite{bertoni1}, \cite{bertoni2} is needed. 
Overhead of reversible implementations, which also offer invertibility, of these functions need to be assessed. With invertibility, asymmetric encryption is also feasible. We are exploring asymmetric encryption direction.
Another important research direction is quantification of security offered by BS-PUF vs AES.

\tim{
The impact of PUF noise requires more discussion.
The proposed design uses raw PUF responses;
it will therefore be noisier than traditional PUFs.
An error coding scheme using helper data and some form of fuzzy extraction is required.
}

\textcolor{red}{
Once we have designs for a realistic permutation family, similar evaluations are needed for their robustness. Path delay distributions across chips need to show variability \textcolor{red}{and uniqueness}; 
within the same PUF different paths need to show variability \textcolor{red}{and randomness}; temperature and supply voltage caused delay variation needs to be small enough. 
In addition, resource needs for these implementations need to be evaluated in terms of area, time and energy. The timer for input capture may impose an insignificant overhead. 
Its accuracy plays a central role in feasibility of BS-PUFs.



\section{Conclusions}
\label{sec:conclusion}
In this work, we explore variety of encryption protocols based on commutative PUFs and propose a circuit implementation of the required commutative PUFs (BS-PUF). Commutativity relies on symmetric delays in forward and backward paths regardless
of the message bit state.
Spectre Monte Carlo simulations indicate only less than 1 bit delay offset is introduced by plaintext 
bit state variation. 
This ensure the commutativity of the system. 
Simulation shows that inter-chip variability (up to $\pm25\%$ chip-to-chip variation) is acceptable. 
These encryption PUFs have potential to root the encryption in hardware, hence increasing robustness beyond current software only solutions.

\tim{
Asymmetric encryption methods are valued for their ability to establish a secure communication channel in 
the absence of  a priori shared secret.
Such methods require complex computations resulting in low throughput compared to symmetric encryption.
%
%
BS-PUF has the potential to provide an asymmetric encryption method with performance similar to AES (symmetric encryption).

Basing encryption in hardware limits the attack surface.
An adversary cannot retrieve the message even when both encryption key and ciphertext are known;
information about the PUF behavior is not available to them.
The behavior of the encryption function becomes a secret.
Thus, more entropy is added to the system. Besides, BS-PUF based encryption provides much better speed and area performance than AES.
}




\bibliographystyle{IEEEtran}
\bibliography{pufs}

\end{document}